# When Should AI Follow? Task Structure and Joint Adaptation by Human and AI Agents[1]

**Prothit Sen**
Assistant Professor of Strategy
Indian School of Business (ISB)
Gachibowli, Hyderabad, India 500111
Tel: +91 9910073788
Email: Prothit_Sen@isb.edu

**Sai Mihir Jakkaraju**
Independent Researcher
Bengaluru, Karnataka, India 560086
Tel: +91 9379999739
Email: saimihirj.j@gmail.com

## Abstract

How should organizations divide and sequence decision tasks between human and artificial agents? We develop a computational model of joint sequential adaptation in which two agents differ in a single, precisely specified way: the memory regime governing how past decisions shape subsequent ones. A *recency-weighted* regime-motivated by behavioral evidence on human adaptation-privileges recent outcomes; a *uniform-memory* regime-motivated by the scale-free consistency of algorithmic updating-weights a window of past outcomes equally. Situated in the lineage of NK/NKC models but developed on its own terms as a sequential-adaptation model, the framework varies task scope (*N*), within-task coupling (*K*), and cross-agent coupling (*C*) across modular and sequenced task structures. Three mechanisms organize the results. First, threshold dynamics create absorbing high- and low-payoff regimes, so adaptation compounds whatever it inherits. Second, uniform memory amplifies inherited trajectories-for good and for ill-whereas recency weighting corrects locally but is volatile at scale. Third, memoryless stochastic adaptation, though inferior on average, functions as an escape mechanism when inherited trajectories are poor. Consequently, joint performance is maximized not by the prevalent “AI-first” design, but when scale-free adaptation follows a high-performing human; broad stochastic search instead rescues sequences initiated by low-performing humans. The model and its experimental validation (a separate study) offer organization designers a task-structural contingency logic for human-AI collaboration and caution against universal prescriptions for AI-first deployment.

[1] Working draft by Prothit Sen (ISB) and Sai Mihir Jakkaraju (Independent Researcher).

**Acknowledgments:**

The manuscript has immensely benefited from the feedback of Phanish Puranam, Aseem Kaul, Hart Posen, Thorbjorn Knudsen, Kannan Srikanth, Maciej Workiewicz, and Aks Zaheer. Prothit Sen is further thankful to Sai Mihir Jakkaraju, an independent researcher working as a lead data scientist in an AI-solutions company, for collaborating on this project and bringing together core principles from Computer Science and industry insights from AI-Human ensemble teams that enriched the practical relevance of their modeling choices and model interpretation. AI agent, Claude, was used for improvement in writing, visualization of the data, and in the execution of the experimental validation study. All modeling choices, theoretical insights, and assertions are our own, and so are errors.

## 1. Introduction

As artificial intelligence (AI) takes on increasingly consequential roles in organizational decision-making, managers confront a design problem that organization theory is unusually well positioned to address: how should decision tasks be divided and sequenced between human and artificial agents (Puranam 2021, Raisch and Fomina 2025)? Practice has largely converged on a default answer. In the prevailing "AI-first" template, AI generates a broad set of candidate options, analyses, or drafts, and humans subsequently select, curate, and refine (Csaszar 2025). This template rests on an intuitively appealing division of labour- machine breadth followed by human judgment-but its performance implications have rarely been examined against alternative task architectures. Whether AI should search first, search second, or search in parallel with humans, and how the answer depends on the structure of the task, remain open questions.

These questions matter because a large literature in organization science establishes that the architecture of a decision process-how a problem is decomposed, who decides what, and in what order-can matter as much as the capability of the individual decision makers (Ethiraj and Levinthal 2004, Christensen and Knudsen 2013, Csaszar and Eggers 2013, Puranam et al. 2015). A parallel, task-based literature in the economics of AI reaches a complementary conclusion: whether AI displaces or complements human work is determined at the level of tasks and their allocation, not at the level of occupations or technologies (Acemoglu and Restrepo 2019, Agrawal et al. 2019, Ide and Talamas 2025). What is missing at the intersection of these literatures is a transparent formal apparatus for studying how *sequencing* and *coupling* between human and AI decision tasks shape joint performance when the two agents adapt differently.

We develop such an apparatus. Our model deliberately reduces the human-AI contrast to a single, precisely specified difference in how agents adapt: the *memory regime* through which past decisions influence subsequent ones. One archetype-motivated by robust behavioral evidence that human adaptive search privileges recent and salient feedback (Levinthal and March 1981, Gigerenzer and Gaissmaier 2011, Posen et al. 2018)-applies *recency-weighted* adaptation: recent realized decisions receive greater weight in shaping the next decision. The other archetype-motivated by the observation that algorithmic systems apply their updating rules with a consistency that does not degrade as the problem scales (Nishant et al. 2024, Csaszar 2025)-applies *uniform-memory* adaptation: all decisions within its memory window receive equal weight. We emphasize at the outset what these archetypes are and are not. Both are heuristics: decades of research in computer science establish that heuristic search is central to, not the antithesis of, artificial intelligence (Simon and Newell 1958, Pearl 1984, Russell and Norvig 2010). The contrast we formalize is therefore not "heuristic human versus rule-based machine," but a contrast between two archetypal memory regimes-fast, scale-dependent adaptation versus slow, scale-free adaptation (cf. Levinthal and Schliesmann 2025)-whose association with human and artificial agents we motivate carefully, and whose limits we discuss explicitly. A third variant, *memoryless stochastic* adaptation, in which an agent conditions on no prior states at all, completes the design space; we use it to study untethered machine exploration, and we are explicit that it is not a model of AI "hallucination" as that term is understood in contemporary AI research (Dell'Acqua et al. 2023).

The agents jointly work a generic decision task represented as a sequence of binary sub-decisions. Three structural parameters govern the setting: the scope of each agent's task ($N$), the depth of coupling among consecutive decisions within an agent's task ($K$), and the coupling between the two agents' tasks ($C$), which determines how many realized decisions of an upstream agent are inherited by a downstream agent. This parameterization places the model in the lineage of NK and NKC models of organizational search (Levinthal 1997, Rivkin and Siggelkow 2003, Ganco et al. 2020), and we retain the familiar notation to preserve comparability of the *design questions*. But we are transparent that the model is not a canonical NK model: decisions are realized progressively rather than reconfigured across a combinatorial landscape, the payoff structure is additive with a single global optimum, and $K$ captures the depth of sequential path dependence rather than epistatic interaction. Section 3.5 maps every construct onto its canonical counterpart and states each departure explicitly, and we develop and justify the model on its own terms as a *sequential-adaptation model*. This positioning follows the methodological dictum that the value of a stylized model lies in the generalizable mechanisms it isolates, not in point predictions (Knudsen et al. 2019).

Three mechanisms, all traceable to the model's threshold dynamics, organize our results. First, because each agent converts a weighted average of past decisions into its next decision via a fixed threshold, adaptation exhibits *absorbing regimes*: once a trajectory's running average settles above (below) the threshold, subsequent decisions tend to reproduce success (failure). Adaptation therefore compounds whatever it inherits. Second, the two memory regimes manage this compounding differently. Uniform memory is a *stabilizer*: it damps fluctuations and faithfully amplifies inherited trajectories-for good and for ill. Recency weighting is a *corrector*: it responds quickly to local feedback but becomes volatile, and eventually biased, as scope and coupling grow. Third, memoryless stochastic adaptation-dominated on average-becomes valuable precisely when inheritance is a liability: it is the only regime that can escape an absorbed low-payoff trajectory.

These mechanisms yield a contingency logic for human-AI task design. In modular structures, where agents adapt in parallel, joint payoff is maximized when the uniform-memory (AI-archetype) agent searches over a moderately, not maximally, broader scope than the recency-weighted (human-archetype) agent, and when the latter faces shallow coupling. In sequenced structures, order matters asymmetrically. When AI leads, and the human follows, gains attenuate as cross-agent coupling deepens, because recency-weighted adaptation degrades into recency bias when forced to absorb many inherited states. When the human leads and AI follows, joint payoff reaches its maximum across the entire parameter space-provided the upstream human trajectory is a high-performing one-because uniform memory compounds high-quality inheritance. When the upstream trajectory is poor, memoryless stochastic adaptation outperforms uniform memory downstream, because escape dominates amplification. On aggregate, a sequence in which AI follows a high-performing human outperforms the AI-first sequence-a direct challenge to the prevailing design template.

Our contribution is threefold. First, we provide a transparent formal contrast between two archetypal adaptation regimes relevant to human-AI task design, specified so that every result can be traced to an explicit mechanism-including results, such as the amplification of high-quality upstream search by uniform memory, that follow closely from the updating rule itself; we treat such traceability as the point of the exercise rather than an embarrassment to it (Knudsen et al. 2019). Second, we establish sequencing and cross-agent coupling as first-order design levers for human-AI collaboration and identify the conditions under which "AI-follows" dominates "AI-first." Third, we reinterpret broad, memoryless machine exploration: rather than a pathology, it is a systematic escape mechanism whose value is contingent on the quality of what it would otherwise inherit. Throughout, we bound our claims deliberately: the model examines a stylized contrast between archetypal search logics and derives implications for human-AI task structure; it is not a comprehensive theory of human-AI complementarity, and we devote an extended section to its boundary conditions.

Finally, we step beyond the simulation toward empirical grounding. Using an AI-assisted experiment in the domain of corporate M&A, predicting announcement abnormal returns across time, we demonstrate the model's central ordering in a real analytical workflow: when the human collaborator brings genuine domain expertise, an expert-first architecture, in which the human curates the variables and the machine fine-tunes the functional form, outperforms a brute-force machine-first architecture that searches broadly but lacks contextual calibration, and whose surplus breadth manifests, observably, as overfitting that fails to transfer from past deals to present ones (Section 6.3; full design and results in the Online Appendix).

## 2. Theoretical Background

### 2.1. From Tools to Agents: Task Structure in Human-AI Collaboration

Early management research on AI predominantly cast the technology as a tool-an aid to prediction and task execution whose organizational consequences run through automation and augmentation of human work (Raisch and Krakowski 2021, Choudhury et al. 2020, Csaszar et al. 2024, Choudhary et al. 2025). Valuable as it is, the tool framing directs attention to whether AI replaces or assists a human on a given activity, and away from a prior design question: how a joint decision process should be architected when the machine itself behaves as a decision-making participant. A growing body of work recognizes precisely this shift-AI systems that generate options, take decisions, and act with meaningful autonomy are better analyzed as agents embedded in an organizational division of labor (Krakowski et al. 2023, Vanneste and Puranam 2024, Raisch and Fomina 2025). From this agentic vantage point, human-AI collaboration becomes an organization design problem (Puranam 2021): performance depends on how decision tasks are partitioned, coupled, and sequenced across heterogeneous agents (Christensen and Knudsen 2013, Puranam et al. 2015).

This organizational lens converges with a task-based tradition in the economics of AI. Acemoglu and Restrepo (2019) locate the displacement-versus-complementarity question at the level of tasks; Agrawal et al. (2019) decompose decision workflows into prediction and judgment components allocated across humans and machines; and Ide and Talamas (2025) show, in a knowledge-hierarchy model, that the organizational position in which autonomous AI is deployed determines which workers it displaces and which it complements. These literatures agree that the consequences of AI are determined by task allocation, yet they typically hold the *process of adaptation* fixed: agents differ in cost or knowledge, not in how their search behavior evolves over the course of a task. Conversely, the behavioral tradition in organization science has developed rich models of adaptive search (Levinthal 1997, Rivkin and Siggelkow 2003, Baumann and Siggelkow 2013, Billinger et al. 2021) but has only begun to incorporate artificial agents whose adaptation differs qualitatively from that of humans (Raisch and Fomina 2025, Csaszar 2025). Our model is aimed at this intersection: it holds the task constant and varies (a) the adaptation regimes of the two agents and (b) the structure-modular versus sequenced, loosely versus tightly coupled-that connects their tasks.

Two scope conditions follow from this positioning, and we flag them now because they discipline the claims we make later. First, following Christensen and Knudsen (2013) and Puranam (2021), we distinguish the *nature of interdependence* (whether agents' decisions relate in parallel or in sequence, and how tightly) from the *nature of specialization* (whether agents perform different kinds of tasks). Our model varies interdependence and sequencing; it does not model domain specialization, and none of our implications concerns it. Second, our outcome construct is the joint performance of a task architecture. Because the aggregate payoff in our model is additive across agents (Section 3), we adopt a deliberately conservative definition of complementarity: a *mixed-regime* configuration (one agent per regime) is complementary when it outperforms the corresponding *pure-regime* configurations on the same task structure and parameters. This definition cannot capture super-additive synergies, and we return to this limitation explicitly in Section 6.4.

## 2.2. Two Archetypal Adaptation Regimes

What, analytically, distinguishes an artificial from a human decision-making agent? An earlier draft of this research framed the contrast as "heuristic (human) versus rule-based (AI)" search. That framing is untenable. Heuristics are foundational to artificial intelligence: Simon and Newell (1958, p. 7) described intelligent machines as using "processes that are closely parallel to human problem-solving processes," and Pearl (1984) formalized heuristics as the core of machine search; the tradition remains central in contemporary AI (Russell and Norvig 2010). Nor is "unbounded rationality" a satisfactory differentiator: AI systems are computationally bounded, exhibit a jagged capability frontier-expert-level on some tasks and mediocre on adjacent ones (Dell'Acqua et al. 2023)-and an omniscient-AI assumption would in any case be observationally equivalent to an expert human, generating no distinctive theory (Csaszar 2025).

We therefore locate the contrast at a narrower and more defensible juncture: the *memory regime* of adaptation-how past realized decisions are weighted when generating the next one-and its *stability under scale*. On the human side, a large body of behavioral evidence indicates that adaptive search weights recent and salient outcomes disproportionately: adaptation is experiential and local (Levinthal and March 1981), governed by aspiration-relative feedback (Cyert and March 1963, Greve 2023, Posen et al. 2018), and implemented through fast-and-frugal heuristics whose ecological rationality depends on selective, recency- and salience-weighted use of information (Gigerenzer and Goldstein 1996, Gigerenzer and Gaissmaier 2011, Artinger et al. 2022). Crucially, this weighting is *scale-dependent*: as task scope and coupling grow, cognitive and attentional limits force even highly capable humans to rely more heavily on recent, available information (Simon 2000, Todd and Gigerenzer 2007). Recency weighting is thus not a claim that humans are simple; it is a parsimonious formal signature of how boundedly rational adaptation degrades under load. It corresponds to what Levinthal and Schliesmann (2025) characterize as *fast learning*: responsive to new feedback, at the price of volatility and potential bias.

On the machine side, we do not claim that contemporary AI systems literally update by equal-weight averaging-they do not, and we make no such claim (see Section 6.4). The property we abstract is *consistency at scale*: an algorithmic agent applies whatever updating rule it embodies uniformly across a decision sequence, without the endogenous drift toward recency that load induces in humans (Nishant et al. 2024, Shick et al. 2024). Uniform weighting over a memory window is the simplest formal representation of this scale-free consistency, and it corresponds to *slow learning* in the sense of Levinthal and Schliesmann (2025): stable, history-preserving, and-as we will show-prone to faithfully amplifying whatever history it inherits. The two regimes are therefore best read as poles of a fast-slow learning continuum, with the association to "human" and "AI" resting on the scale-(in)dependence argument above. A skeptical reader may prefer to interpret our model as contrasting more and less recency-biased searchers in general; our structural results about sequencing and coupling survive that reading, though the motivating application to human-AI task design is what gives the parameter ranges (e.g., the machine agent's much larger scope) their substantive interpretation.[2]

---

[2] A further asymmetry that the archetypes do not capture is the jagged capability frontier of contemporary AI (Dell'Acqua et al. 2023)-region-dependent reliability across a task-which we discuss as an extension in Section 6.4.

Alongside memory weighting, the archetypes differ in *scope*. Machine agents can evaluate and act over decision sequences orders of magnitude longer than human working capacity permits (Raisch and Fomina 2025). We implement this as a scope asymmetry-the machine agent's task contains more sub-decisions than the human agent's-and study how joint performance varies with the *ratio* of scopes rather than fixing an arbitrary level. Finally, we consider a third regime, *memoryless stochastic adaptation*, in which the agent conditions on no history and realizes each decision at random. We introduce it for two reasons: it completes the memory-regime design space (full window weighting, recency weighting, no memory), and it operationalizes a managerially salient scenario-machine exploration untethered from inherited context. We stress the label: this is *not* a model of AI hallucination, which in contemporary usage denotes confident, non-random error arising from overgeneralization or distributional mismatch, not unbiased sampling. We use "memoryless stochastic adaptation" throughout and discuss the distinction in Section 6.4.

### 2.3. Task Structure: Modularity, Sequencing, and Coupling

The second building block is the structure connecting the two agents' tasks. Organization design research distinguishes architectures in which decision tasks proceed in parallel with pooled outcomes from architectures in which one agent's output becomes another's input (Thompson 1967, Ethiraj and Levinthal 2004, Christensen and Knudsen 2013). We examine three: a *modular* structure, in which the human-archetype and machine-archetype agents adapt independently and outcomes are pooled; and two *sequenced* structures, AI→H and H→AI, in which the downstream agent inherits a portion of the upstream agent's realized decisions as the starting history for its own adaptation. The intensity of inheritance-how many upstream decisions the downstream agent must build upon-is governed by a cross-agent coupling parameter, *C*, in the spirit of the C parameter in NKC models of coupled search (Kauffman 1993, Ganco et al. 2020), though implemented here as inherited history rather than cross-landscape epistasis (Section 3.5).

Sequencing interacts with memory regimes in ways that are difficult to intuit ex ante, which is what motivates a computational treatment (Harrison et al. 2007, Puranam et al. 2015). A downstream agent's adaptation is path-dependent on inherited states; how it metabolizes that inheritance depends on its memory regime. A recency-weighted agent discounts inherited history quickly-correcting bad inheritance but also squandering good inheritance. A uniform-memory agent preserves inherited history-compounding good inheritance but also entrenching bad inheritance. A memoryless agent ignores inheritance altogether. Whether each property is a strength or a liability depends on the coupling depth, the relative scopes, and the quality of what is inherited. The model formalizes these interactions and traces their joint-performance consequences.

## 3. The Model

### 3.1. Design Philosophy and Scope

Our model is intentionally minimal. Following the methodological position articulated by Knudsen et al. (2019), we treat the model not as a forecasting device but as an instrument for isolating mechanisms: every result reported below is accompanied by an account of *why* it arises from the specified updating rules and task structure, and results that follow closely from an updating rule are presented as such. Three commitments discipline the exposition. First, *construct honesty*: each parameter is named for what it does in the model, and Table 1 maps each construct onto its canonical NK/NKC counterpart, stating the departures. Second, *incrementality*: we establish baseline behavior for each adaptation regime in isolation (Section 3.6, Figure 1) before introducing two-agent structures, so that readers can attribute each subsequent result to the added element. Third, *counterfactual grounding*: mixed-regime results are evaluated against pure-regime baselines wherever the model's structure permits exact comparison (Section 3.6), and we state plainly where it does not.

---------- Insert Table 1 about here ----------

**3.2. Task Representation and Payoff**

A task assigned to agent $a \in \{H, AI\}$ is a sequence of $N_a$ binary sub-decisions, $x_1^a$, $x_2^a$, …, $x_N^a$, realized one at a time in order. Each sub-decision resolves to a state of 1 (an adequate resolution, contributing positively to task performance) or 0 (an inadequate resolution, contributing nothing). The intended empirical referent is the broad class of organizational tasks that unfold as ordered chains of quality-gradable sub-decisions: successive analytical steps in a research or diligence workflow, sequential engineering change decisions, stage-wise underwriting or triage judgments, or the stepwise construction of a strategic recommendation. In such settings, performance is well approximated by the proportion of steps resolved adequately, and each step is taken against the backdrop of the steps already taken. Formally, agent *a*'s realized payoff is the share of its sub-decisions resolved to 1:

$$PO^a = (1/N_a)\ \Sigma_{i=1..Na}\ x_i^a\ , \qquad a \in \{H, AI\}.$$

The joint payoff of a two-agent configuration is the average of the two agents' payoffs, $APO = (PO^H + PO^{AI})/2$, computed over 1,000 simulation runs per parameter combination (Clement and Puranam 2018). Two properties of this payoff structure should be stated at the outset rather than discovered by the reader. First, it is additive: the all-ones configuration is the unique global optimum, and the payoff surface over configurations contains no interactions among decision states. Whatever ruggedness-like phenomena arise in this model arise from the *dynamics of* *adaptation* (Section 3.6), not from the payoff function. Second, because APO is an average, it is bounded above by the better agent's payoff; it therefore measures the performance of the task architecture, and our conservative definition of complementarity (Section 2.1) compares mixed-regime architectures against pure-regime architectures rather than asserting super-additivity.[3]

### 3.3. Adaptation Regimes

All agents adapt by the same template: the next decision is generated by comparing a weighted mean of the $K$ most recently realized decision states against a threshold of 0.5. The regimes differ only in the weights. Initialization is symmetric: the first $K$ states of an agent that starts a task without inheritance are drawn i.i.d. Bernoulli(0.5). Threshold ties resolve to 1 for all agents, introducing a slight upward drift that is common across regimes and therefore does not favor any agent.[4]

*Recency-weighted adaptation (human archetype).* Let the $K_H$ most recent realized states, ordered from oldest to most recent, be $x^{H}_{i-K_n+1}, \ldots, x_i^{H}$. The weight on the $j$-th of these states (j = 1 oldest, j = $K_H$ most recent) is $w_j = j$, so weights increase linearly with recency. The next decision is:

$$x_{i+1}^{H} = 1 \text{ if } [\, \Sigma_{j=1..K_n} j \cdot x^{H}_{i-K_n+j} \,] \,/\, [\, \Sigma_{j=1..K_n} j \,] \geq 0.5 \,; \quad x_{i+1}^{H} = 0 \text{ otherwise.}$$

We name this construct for what it is-recency-weighted adaptation-rather than "satisficing." Satisficing in its canonical sense involves an aspiration level and a stopping rule

[3] APO lies in [0, 1]. Because initialization is symmetric Bernoulli(0.5), values cluster around 0.5; all comparisons reported in this paper are within-model comparisons across task architectures at identical parameters, and no cardinal interpretation of payoff levels is intended.

[4] An earlier version of this work contained an inconsistency between the threshold equation and a worked example on the resolution of ties; we standardize here on ties resolving to 1 in all equations, examples, and code. The induced drift is identical across regimes and thus affects levels, not comparisons.

(Simon 1955, Artinger et al. 2022); our rule involves neither. The behavioral motivation, developed in Section 2.2, is that recency weighting is the parsimonious signature of experiential, feedback-driven human adaptation and of the drift toward recent information that load induces (Levinthal and March 1981, Gigerenzer and Gaissmaier 2011). The linear weighting scheme is one of many possible; it suffices to establish the qualitative contrast, and Section 6.4 discusses alternatives.[5]

*Uniform-memory adaptation (AI archetype).* The machine-archetype agent applies equal weights over the same kind of window:

$$x_{i+1}^{AI} = 1 \text{ if } (1/K_{AI}) \Sigma_{j=1..K_{ai}} x^{AI}_{i-K_{ai}+j} \geq 0.5 \,; \quad x_{i+1}^{AI} = 0 \text{ otherwise.}$$

Within its window the agent preserves history without discounting-the scale-free consistency property motivated in Section 2.2. We correct here an ambiguity in an earlier version of this work: conditioning is on the window of the $K_{AI}$ most recent states, not on all prior states; when $K_{AI}$ is large relative to the realized sequence the two coincide. Figure 2 presents a worked example in which the two regimes, facing an identical realized history, reach different decisions-the recency-weighted agent extrapolates the recent run of failures, while the uniform-memory agent preserves the older successes.

---------- Insert Figure 2 about here ----------

*Memoryless stochastic adaptation.* The third regime conditions on no history:

$$x_{i+1} \quad \text{Bernoulli(0.5), independently of all prior states.}$$

[5]Any weighting scheme that is strictly increasing in recency (e.g., exponential discounting of older states) preserves the qualitative contrast with uniform weighting; the linear scheme is the simplest member of that class.

This regime operationalizes untethered exploration-search that neither preserves nor discounts inherited context but ignores it. As emphasized in Section 2.2, it is not a representation of AI hallucination; hallucination in contemporary AI research denotes systematic, confidence-preserving error, not unbiased noise. We deploy memoryless adaptation in one specific role: as an alternative downstream regime in H→AI sequences (Section 3.4), where its analytical function-severing path dependence-is exactly what the design question requires.

### 3.4. Structural Parameters: Scope, Coupling Depth, and Cross-Agent Coupling

*Task scope (N).* $N_a$ is the number of sub-decisions in agent *a*'s task-the length of its decision sequence. Motivated by the scope asymmetry of Section 2.2, we impose $N_{AI} > N_H$ and treat the scope ratio $N_{AI}/N_H$ as a primary design variable, sweeping it from near parity to more than an order of magnitude.

*Coupling depth (K).* $K_a$ is the number of preceding realized decisions that condition agent *a*'s next decision-the window length in the adaptation rules above. Substantively, K captures the depth of sequential coupling within a task: how far back the resolution of earlier sub-decisions reaches into the determination of the current one. High-K tasks are those in which each step must be taken in light of a long stretch of prior commitments (e.g., late-stage engineering changes constrained by many earlier design decisions); low-K tasks are those in which each step depends only on its immediate predecessors. Two clarifications are essential. First, K in this model is *not* the epistatic interdependence parameter of canonical NK models: it does not enter the payoff function and does not generate a multi-peaked payoff surface. It is a property of the *decision process*-the depth of path dependence in adaptation. Second, because K parameterizes the window of the agent working the task, task and agent are not separable in this construct: assigning agent *a* task of coupling depth $K_a$ means that agent *a*'s adaptation conditions on $K_a$ prior states. We therefore interpret $K_H$ and $K_{AI}$ as properties of the human-assigned and machine-assigned *task components*, and we confine interpretations of K to this process sense throughout.[6]

*Cross-agent coupling (C).* In modular structures, C = 0: the agents adapt independently and outcomes are pooled. In sequenced structures, C > 0: after the upstream agent completes its task, its first C realized decision states are inherited by the downstream agent as the initial history on which the downstream agent's own adaptation rule operates. In AI→H, the human inherits $S^H = (x_1^{AI}, \ldots, x_C^{AI})$ and continues with recency-weighted adaptation; in H→AI, the machine inherits $S^{AI} = (x_1^H, \ldots, x_C^H)$ and continues with uniform-memory adaptation. C thus measures the intensity of sequential interdependence between the two task components: larger C means the downstream agent's trajectory is seeded by-and its early decisions computed from-a longer stretch of the upstream agent's realized search. This implementation is inspired by, but simpler than, the cross-landscape epistasis of NKC models (Kauffman 1993, Ganco et al. 2020): inherited history rather than coupled payoff functions (Table 1).

One asymmetry of the H→AI structure deserves emphasis because it drives a key result. Since $N_{AI} > N_H$, the downstream machine agent eventually exhausts the human's realized states: once its rolling window has moved past the inherited segment, it conditions only on its own realized states. The "memory" of upstream human search therefore fades endogenously as the machine's scope grows-fast when $N_{AI}/N_H$ is large, slowly when scopes are comparable. Table 2 presents a worked example of this inheritance-and-fade process under uniform-memory adaptation; the memoryless variant simply replaces the post-inheritance rule with Bernoulli(0.5) draws.

[6]We acknowledge the standard point that in canonical NK models complexity is a property of the environment, not the agent. Because our K is a process construct in which task assignment and adaptation window coincide, we avoid the term "complexity" for it and speak of coupling depth throughout.

---------- Insert Table 2 about here ----------

### 3.5. Relationship to Canonical NK/NKC Models

Because the model borrows NK/NKC notation, we state precisely where it coincides with and departs from the canonical apparatus; Table 1 summarizes. (i) *Payoff structure.* Canonical NK models define fitness through contribution functions in which each decision's payoff depends on K others, generating rugged, multi-peaked landscapes (Levinthal 1997). Here the payoff is additive with a unique global optimum (all ones); there is no epistasis in the payoff function. (ii) *Search process.* Canonical NK agents move between complete configurations-evaluating neighbors and relocating on a landscape. Here agents realize decisions progressively and never revisit them: search is the sequential narrowing of an incomplete configuration to a complete one. Consequently, the canonical local/distant search distinction has no direct counterpart. (iii) *K.* Canonical K indexes epistatic interdependence in the environment. Here K indexes the depth of the adaptation window-sequential coupling and path dependence in the decision process. (iv) *C.* Canonical C couples payoff functions across landscapes. Here C is the length of inherited history across agents. (v) *Local traps.* The canonical notion of a local peak-a configuration whose neighbors are all inferior-does not exist here. Its functional analogue is an *absorbing low-payoff regime* of the adaptation dynamics: a trajectory whose window average has fallen below the threshold and therefore reproduces failures (Section 3.6). When we use trap-like language below, it refers to this precisely defined dynamic phenomenon, not to landscape topology.

Why, then, retain the N-K-C notation at all? Because the *design questions* the model addresses-how scope, within-task coupling, and cross-agent coupling shape the performance of alternative divisions of adaptive labor-are the questions the NK/NKC tradition taught the field to ask (Ethiraj and Levinthal 2004, Rivkin and Siggelkow 2003, Ganco et al. 2020), and preserving the notation makes the correspondence of questions (and the non-correspondence of mechanics) easy to audit. Readers who prefer to see the model as a bespoke sequential-adaptation model are reading it exactly as intended.

### 3.6. Baseline Behavior: Threshold Dynamics and Absorbing Regimes

Before introducing two-agent structures, we characterize how each regime behaves alone-both to establish the mechanism that drives all subsequent results and to create pure-regime counterfactuals. The engine of the model is a threshold dynamic. Each next decision reproduces the sign of the window average relative to 0.5. A trajectory whose window average sits above the threshold tends to generate further 1s, which keep the average high: a self-reinforcing *high-payoff regime*. Symmetrically, a window average below the threshold generates 0s that keep it low: an *absorbing low-payoff regime*. Early stochastic realizations therefore matter disproportionately-they determine which regime a trajectory enters-and the memory weights determine how easily a trajectory, once absorbed, can exit. Uniform weights damp fluctuations: a long run of contrary outcomes is required to move the average across the threshold, so the uniform-memory regime is stable in both directions-it locks in success and locks in failure. Recency weights amplify the most recent outcomes: short runs can flip the trajectory, so the recency regime exits low-payoff regimes more easily but also falls out of high-payoff regimes more easily. This stability-responsiveness trade-off is the model's core mechanism, and it is the sequential-adaptation analogue of the exploration-exploitation and fast-slow learning trade-offs of the behavioral tradition (March 1991, Levinthal and Schliesmann 2025).

Figure 1 reports baseline payoffs for each regime alone across the (N, K) grid: panel (a) for uniform-memory adaptation, panel (b) for recency-weighted adaptation. Three features matter. First, under uniform memory, payoff rises with K at low K and the gains are convex in scope: longer windows damp the volatility of the running average, and-because symmetric Bernoulli initialization plus the tie-to-1 rule impart a slight upward drift-damping stabilizes trajectories disproportionately in the high-payoff regime. This is the mechanism behind the baseline pattern, stated here so that the two-agent results can be read against it. Second, under recency weighting the surface tilts the opposite way: performance is strongest where K is low-short windows keep volatility moderate and responsiveness valuable-and degrades as K rises, because with long windows the recency weights concentrate effective memory in a few recent states, discarding the information in the rest of the window. At high K, recency weighting behaves as *recency bias*. Third, the crossing pattern between the panels establishes that neither regime dominates: which regime performs better is itself a function of task structure. These single-regime surfaces also serve as exact pure-regime counterfactuals for the modular analysis: because modular joint payoff is the average of two independent single-agent payoffs, the expected APO of a pure H-H (or AI-AI) modular pair at given parameters equals the corresponding single-agent baseline in Figure 1. Mixed-regime modular results below can therefore be benchmarked directly against same-task pure-regime configurations. For sequenced structures no such exact reduction exists, and we note where pure-regime sequenced counterfactuals remain future work (Section 6.4).

---------- Insert Figure 1 about here ----------

*Simulation protocol.* For each parameter combination ($N_H$, $N_{AI}$, $K_H$, $K_{AI}$, C, task structure, downstream regime), we simulate 1,000 independent runs and report mean payoffs. Scope ratios $N_{AI}/N_H$ range from just above 1 to beyond 15; coupling depths $K_a$ range from 1 to $N_a - 1$; C ranges from 1 to below N of the upstream agent. Reported surfaces are smoothed with cubic splines over the 1,000-run cell means; raw sensitivity grids are reported unsmoothed in Section 5.

## 4. Results

### 4.1. Modular Task Structures (C = 0)

Figure 3 reports joint payoff under modular structures as a function of the scope ratio $N_{AI}/N_H$, for varying relative coupling depths $K_H/K_{AI}$. Two patterns organize the results. First, joint payoff is curvilinear-inverted U-shaped-in the scope ratio: performance declines both when the machine agent's scope is close to the human's and when it is very large, peaking at moderate asymmetry ($N_{AI}/N_H \approx 5$). Second, the peak is highest when the human-archetype agent's task is shallowly coupled ($K_H/K_{AI} < 0.5$); assigning the recency-weighted agent a low-K task component is the single most reliable lever in the modular design space.

*Mechanism.* Both patterns follow from the threshold dynamics of Section 3.6. The machine agent's contribution to APO rises with scope only insofar as longer sequences give its damped, slightly upward-drifting trajectory more room to consolidate in the high-payoff regime; but a longer sequence also raises the cumulative probability that an unlucky early window absorbs the trajectory in the low-payoff regime, where uniform memory's stability becomes a liability-it stays there. Moderate scope asymmetry balances consolidation against absorption risk. On the human side, low K keeps recency weighting in the range where it is responsiveness rather than bias: with short windows, the recency-weighted agent exits low-payoff regimes quickly, and its volatility costs little. As $K_H$ rises, effective memory concentrates on the last few states and the agent's trajectory whipsaws, dragging the average down. Benchmarked against the pure-regime counterfactuals of Figure 1, the mixed configuration in this region outperforms both the all-uniform and all-recency modular pairs on the same task parameters-complementarity in the conservative sense of Section 2.1: each regime is deployed where the other's weakness would bind.

*Proposition 1a. In modular task structures, joint payoff is curvilinear (inverted U-shaped) in the relative scope of machine to human search ($N_{AI}/N_H$) and is maximized when moderate scope asymmetry is combined with shallow coupling depth in the human-assigned task component (low $K_H/K_{AI}$).*

The aggregate pattern notwithstanding, high joint payoffs also arise when the human-assigned component is deeply coupled ($K_H/K_{AI} > 2$)-but only at the two extremes of the scope ratio. At very low $N_{AI}/N_H$, the mechanism is *containment*: with a short machine sequence, absorption risk on the machine side is limited, so the volatile human trajectory is paired with a small but reliable machine contribution, and the average holds up. At very high $N_{AI}/N_H$, the mechanism is *dilution*: APO weights the two agents equally, but the machine agent's long, consolidated trajectory contributes a stable high payoff that offsets the human component's weakness-sheer scale of the well-suited component compensates for the poorly suited one. These are boundary-condition designs, not first-best ones: both lie below the global peak of Proposition 1a, and the dilution design in particular buys robustness with a large machine deployment.

*Proposition 1b. In modular task structures with deep coupling in the human-assigned component (high $K_H/K_{AI}$), joint payoff exhibits two local maxima in relative scope: one at very low and one at very high $N_{AI}/N_H$.*

---------- Insert Figure 3 about here ----------

### 4.2. Sequenced Structure: AI→H (C > 0)

In the AI→H structure the machine agent completes its task first; the human agent inherits the machine's first C realized states as initial history and continues with recency-weighted adaptation (Figure 4). The modular result survives qualitatively-joint payoff remains curvilinear in the scope ratio-but two attenuation effects now bind. First, gains fall as cross-agent coupling deepens: performance deteriorates markedly once C exceeds roughly twice the machine's coupling depth ($C/K_{AI} > 2$). Second, attenuation is amplified when the machine's scope is large ($N_{AI}/N_H > 5$).

*Mechanism.* Both effects reflect what recency weighting does to inherited history. The inherited segment functions as the initial window of the human's adaptation; with recency weights, the informational content of that inheritance is concentrated in its last few states, and the remainder is effectively discarded. When C is small, this is calibration: the human conditions on a short, digestible summary of the machine's search and adapts responsively from there. When C is large, it is recency bias: the design forces a long stretch of upstream context onto an agent whose regime structurally cannot retain it, so most of the inherited information is lost precisely when the task structure makes it matter most. The scope amplification compounds this: a long upstream machine sequence has typically consolidated deep in one regime, so the states the human inherits are highly autocorrelated; a recency-weighted window over autocorrelated states is volatile at the regime boundary, and downstream adaptation adds noise rather than refinement. In design terms, positioning a recency-weighted agent downstream is advisable only under loose cross-agent coupling; tightening the coupling converts the human's responsiveness into bias.

> *Proposition 2. In AI→H sequenced structures, joint gains attenuate as cross-agent coupling (C) deepens relative to the machine's coupling depth, and the attenuation is amplified by the machine's relative scope ($N_{AI}/N_H$).*

---------- Insert Figure 4 about here ----------

**4.3. Sequenced Structure: H→AI (C > 0)**

In the H→AI structure the human agent completes its (narrower) task first; the machine inherits the first C realized human states and continues, either with uniform-memory adaptation or with memoryless stochastic adaptation once inheritance is exhausted (Section 3.4, Table 2). Figure 5 reports the results, which we organize by the quality of the upstream trajectory-operationalized as the human's realized first-stage payoff.

When the upstream human trajectory is high-performing, downstream uniform-memory adaptation produces the highest joint payoffs observed anywhere in the parameter space, and the advantage is strongest under deep cross-agent coupling and persists even at large downstream scope ($N_{AI}/N_H > 15$). We are deliberate about the status of this result. Its core is a direct consequence of the uniform-memory rule: an agent that averages inherited states without discounting will, when seeded with a predominantly-1 history, generate further 1s-the compounding is anchored in the arithmetic of the updating rule, as a careful reader will note. We regard this not as an artifact to be explained away but as the mechanism itself, formally exposed: slow, history-preserving adaptation is an *amplifier of inheritance* (Levinthal and Schliesmann 2025). What the simulation adds beyond the arithmetic is the boundary structure of the amplification, which is not derivable by inspection: (i) the advantage requires sufficient coupling-at low C the inherited seed is too short to anchor the downstream window, and performance reverts toward the machine baseline; (ii) the advantage erodes with scope through the endogenous memory-fade of Section 3.4-as $N_{AI}/N_H$ grows, the machine's window rolls past the inheritance and the trajectory's fate is increasingly determined by its own realized states, so the marginal value of upstream quality declines; and (iii) the interaction of (i) and (ii) defines a ridge in (C, $N_{AI}/N_H$) space along which amplification is maximized. The design lesson is correspondingly precise: sequencing a uniform-memory agent downstream converts upstream human quality into system performance, at a rate governed by coupling and dissipated by unbounded scope.

> *Proposition 3a. In H→AI sequenced structures, joint payoff is maximized when downstream uniform-memory adaptation inherits a high-performing upstream human trajectory under deep cross-agent coupling; the advantage attenuates as downstream scope grows and inherited memory fades.*

When the upstream human trajectory is low-performing, the ranking of downstream regimes reverses: memoryless stochastic adaptation modestly but consistently outperforms uniform-memory adaptation, and the advantage persists as downstream scope grows. The mechanism is the mirror image of Proposition 3a. A low-performing upstream trajectory seeds the downstream window below the threshold; uniform memory, faithful to its inheritance, reproduces the failure-the absorbing low-payoff regime of Section 3.6 propagates across the agent boundary. Memoryless adaptation severs the path dependence: each draw is unconditioned, so the downstream trajectory samples the space at rates independent of the inherited regime, and over a long sequence the expected share of adequate resolutions reverts toward the unconditional mean rather than remaining anchored below it. This is a known property of random search-noise helps when search is stuck-and we do not present its direction as novel. What is organizationally meaningful is its locus and scale: in human-only settings, stochastic exploration at this breadth is infeasible (cognitive fatigue, attentional lock-in, coordination costs), so escape-by-randomization has not been an available design lever; a machine agent can implement it over sequences orders of magnitude longer than the inherited segment, making randomization-at-scale a deployable remedy specifically for architectures whose upstream human component is weak. The prescription is conditional pragmatism: where upstream human capability is strong, invest in history-preserving downstream refinement; where it is weak, breadth of untethered exploration dominates sophistication of refinement.

*Proposition 3b. In H→AI sequenced structures with a low-performing upstream human trajectory, downstream memoryless stochastic adaptation yields higher joint payoff than downstream uniform-memory adaptation, and the advantage persists at large downstream scope.*

---------- Insert Figure 5 about here ----------

### 4.4. Comparing Sequences: When Should AI Follow?

The two sequenced structures can now be compared directly. Figure 6 shows that the H→AI structure seeded by a high-performing upstream human dominates the AI→H structure across the examined parameter space: the average joint payoff of the former exceeds even the maximum of the latter. Figure 7 organizes the comparison into three canonical cases, and Figure 8 quantifies the mechanism that separates them.

*Case 1 (AI upstream, high-performing human downstream).* The machine's broad upstream search generates a wide set of interim solutions; the downstream human, adapting under recency weights, conditions on-and therefore utilizes-only a thin recent slice of them. Figure 8 quantifies the resulting *search wastage*: we plot final payoff against the number of interim solutions generated upstream and utilized downstream, where an interim solution is operationally defined as a realized subsequence whose running payoff exceeds that of the adjacent realized alternatives during search (the figure labels these "local peaks"; we note that the term is used in this operational sense, not in the canonical landscape sense-Section 3.5). Even inside the high-payoff band of the scope ratio ($4 < N_{AI}/N_H < 6$), the machine generates over 3.5 times more interim solutions than the human utilizes. The unused surplus is pure architectural loss: computation spent generating context that the downstream regime structurally discards.

*Case 2 (high-performing human upstream, AI downstream).* The upstream human's trajectory begins in the high-payoff regime; the downstream uniform-memory agent inherits and amplifies it (Proposition 3a). Wastage is minimal because the downstream regime retains, rather than discards, inherited search. This is the globally dominant architecture in our results.

*Case 3 (low-performing human upstream, memoryless AI downstream).* The upstream trajectory is absorbed in the low-payoff regime; downstream memoryless breadth provides escape (Proposition 3b). Performance improves on what uniform-memory refinement would deliver but remains below Cases 1 and 2-escape restores optionality, it does not manufacture quality. The expected performance ordering is therefore: Case 2 > Case 1 > Case 3.

A contemporary illustration makes the logic concrete. Consider working with a large language model on an analytical task. An expert-first workflow-the analyst formulates a precise, high-quality specification ("on this data, regress price on lagged demand with individual fixed effects") that the model then extends and refines-instantiates Case 2: the machine's history-preserving elaboration compounds the quality of the upstream specification. A machine-first workflow-soliciting a broad menu ("analyze this price-demand data") from which the analyst selects-instantiates Case 1: much of the generated breadth goes unused, and the analyst's attention gravitates to the most recent or salient candidates (Acar 2023). And the practice of deliberately raising a model's sampling randomness to shake a stalled brainstorm loose from a weak framing instantiates Case 3. Analogous architectures appear wherever organizations sequence expert judgment and machine processing-triage-then-review pipelines in radiology, screening workflows in venture diligence, drafting workflows in legal and policy work-and the model's prescription in each is the same: the value of putting the machine downstream rises with the quality of the human upstream.

---------- Insert Figures 6, 7, and 8 about here ----------

**5. Sensitivity Analyses**

We probe the sensitivity of each result to its governing parameters using unsmoothed grids of cell means (1,000 runs per cell); Figures 9-11 report the grids as heat maps with cell values printed, replacing the smoothed surfaces of Section 4 with the raw quantities.

*Modular structures.* Figure 9 varies relative scope (rows) against relative coupling depth (columns). Joint performance is substantially more sensitive to the coupling-depth ratio than to the scope ratio: within any row, moving the human-assigned component toward shallower coupling shifts payoff more than comparable movements across rows. This is consistent with the mechanism of Proposition 1a-the binding constraint in modular designs is keeping the recency-weighted agent out of the bias regime, not fine-tuning machine scope.

*AI→H structures.* The attenuation logic of Proposition 2 turns on the ratio of within-task coupling to cross-agent coupling faced by the downstream human. Figure 10 varies relative scope against $K_H/C$. As within-task coupling rises relative to cross-agent coupling, joint performance stabilizes: the human's adaptation is increasingly governed by its own task's structure rather than by inherited machine states, and in the limit the structure converges toward the modular case, reproducing the $C = 0$ patterns. Cells in which trajectories failed to converge within the run budget are marked "n.c."

*H→AI structures.* Figure 11 varies relative scope against C separately for the two downstream regimes. Under uniform-memory adaptation (panel a), payoffs are highest and most robust to scope at low-to-moderate coupling ($C = 2$ to $4$), where the inherited window anchors without the memory-fade of very long downstream sequences overwhelming it. Under memoryless adaptation (panel b), the benefits of breadth concentrate at low coupling (C = 2), where weak anchoring to upstream states lets stochastic exploration operate on the largest effective space; as coupling deepens (C > 4), preserving upstream solutions grows in value and uniform memory dominates randomization. The two panels jointly trace the boundary at which the amplifier logic of Proposition 3a gives way to the escape logic of Proposition 3b.

---------- Insert Figures 9, 10, and 11 about here ----------

## 6. Discussion

### 6.1. Theoretical Implications

The paper's first implication concerns the micro-foundation appropriate for theorizing human-AI collaboration. Neither "heuristics versus rules" nor "bounded versus unbounded rationality" survives contact with what is known about machine search (Simon and Newell 1958, Pearl 1984, Dell'Acqua et al. 2023). We propose-and formalize-a narrower foundation that does: the memory regime of adaptation and its stability under scale. This foundation connects the human-AI question directly to the behavioral tradition's fast-slow learning distinction (Levinthal and Schliesmann 2025) and to the stability-responsiveness trade-off at the heart of exploration-exploitation reasoning (March 1991), and it yields testable structure: any pairing of a fast, recency-weighted adapter with a slow, history-preserving adapter should exhibit the sequencing asymmetries we document, whether the slow adapter is an algorithm, a formal procedure, or an organizational routine. The specific application to human-AI task design follows from the scale-dependence argument-humans drift toward recency under load; machine rules do not-but the mechanism is more general than the application, and we regard that generality as a feature the model makes visible rather than a threat to its relevance.

Second, the results establish sequencing and cross-agent coupling as first-order design levers in human-AI collaboration, and they identify the specific prior belief the paper challenges: the presumption, common in both scholarship and practice, that the natural architecture is machine breadth first, human judgment second (Csaszar 2025). In our model that architecture is dominated-not marginally, but by the full gap between the maximum of one design and the average of another-whenever a high-performing human is available to search upstream. The reason is architectural, not motivational: downstream recency-weighted adaptation structurally discards most of what broad upstream machine search generates (the wastage of Figure 8), whereas downstream history-preserving adaptation compounds what upstream human quality provides. This finding complements the task-based economics of AI: where Ide and Talamas (2025) show that an autonomous machine's position in a knowledge hierarchy determines whom it displaces, we show that a machine's position in a decision *sequence* determines whether its defining behavioral property-history-preserving consistency-functions as an amplifier of human quality or as an entrencher of human error. Order of operations, largely undertheorized in the automation-versus-augmentation conversation (Raisch and Krakowski 2021), emerges as a variable of the same rank as allocation itself.

Third, the model reinterprets untethered machine exploration. Random search that escapes entrenched trajectories is a familiar result; our contribution is to locate *when it is the design of choice and why its feasibility is new*. Escape-by-randomization dominates history-preserving refinement precisely when the inherited trajectory is poor, and it becomes a deployable organizational lever only because machine agents can randomize over scales at which human stochastic exploration is infeasible. The practical corollary inverts a common intuition: organizations whose human capital is weakest benefit least from sophisticated, context-preserving AI-the sophistication faithfully preserves the weakness-and most from cheap, broad, weakly tethered exploration.

Finally, the paper carries a methodological implication. Formal work on AI in organizations will often require models that borrow the design questions of an established tradition while departing from its mechanics (Knudsen et al. 2019).

### 6.2. Managerial Implications

Within the model's scope-the structure of interdependence and sequencing, not the content of specialization-the design guidance is specific. First, audit the task architecture before choosing the deployment template: the performance of any human-AI configuration in our results is governed by three auditable properties-the coupling depth of the human-assigned component, the coupling between the human and machine components, and the relative scope of machine search. Second, where strong human expertise exists upstream, prefer the expert-first sequence and couple tightly: machine refinement compounds expert quality, and the gains rise with coupling until downstream scope is so large that inherited context fades. Third, where humans sit downstream of machine search, keep the cross-agent coupling loose: forcing long machine context onto recency-weighted human adaptation converts judgment into recency bias, and much of the machine's search will be discarded regardless-breadth purchased upstream of a human filter is substantially wasted expenditure. Fourth, calibrate AI sophistication to human capability: with weak upstream human capital, broad and weakly tethered machine exploration outperforms context-preserving refinement, so pragmatic investments in scale can dominate investments in sophistication. The common thread is a caution against universal prescriptions-including the currently fashionable AI-first template: in this model there is no dominant deployment pattern, only architectures matched or mismatched to task structure and human capability.

### 6.3. An Experimental Validation: Expert-First versus Brute-Force Machine Search in M&A Prediction

The model's headline ordering, that machine search creates more value when it follows informed human judgment than when it leads with breadth, is a claim about search architectures, and such claims invite demonstration outside the simulation. We therefore conducted a proof-of-concept experiment in a canonical empirical task: predicting cumulative abnormal returns (CAR) around M&A announcements. A dataset of 33,860 deals was split at the median announcement date: the earlier half (2008 to 2013; N = 16,930) is the past context in which models are fitted, and the later half (2014 to 2020; N = 16,930) is the present context in which they must predict. The architectures differ in where, and whether, informed human judgment enters the sequence (Figure 12). *In the expert-first architecture* the human moves first, curating five deal characteristics from M&A theory and choosing three at a time (all ten combinations), and the machine moves second: a penalized learner (LASSO) fine-tunes the functional form within the expert space, searching over the three linear terms and their three pairwise interactions. This instantiates a high-performing upstream human trajectory refined downstream by the machine (Case 2). *In the brute-force architecture* the machine moves first, searching all seven available variables and all 21 pairwise interactions with no upstream curation, and the human moves second in the guise of the regularization filter, a proxy for managers picking the most important surviving variables out of machine breadth for use in the present context (Case 1). A third configuration completes the design space: a novice human who exercises no informed judgment at either end, neither curating variables upstream nor filtering the machine's output downstream, so the full 99-parameter specification is estimated as-is (Case 3). In all arms the selected terms are re-estimated by OLS with industry fixed effects on the past sample, and the fitted model predicts CAR for every deal in the present sample.

---------- Insert Figure 12 about here ----------

The results reproduce the expected performance ordering of the model's three canonical cases, Case 2 above Case 1 above Case 3 (Figure 13). Nine of the ten expert-initiated models transfer to the present context better than the filtered brute-force model, and all ten beat the novice configuration, which attains the best in-sample fit of any model considered and the worst out-of-sample transfer: its surplus breadth is spent fitting past-specific noise that the present context does not reward. Relative to a naive forecast of the past mean, the expert-initiated models pay an average penalty of +0.17 percent, the filtered brute-force model +0.19 percent, and the novice configuration +0.28 percent. Two further features sharpen the correspondence. First, no specification beats the naive forecast, because announcement returns are overwhelmingly idiosyncratic; the expert-first architecture wins by losing less, and the value of expertise is therefore architectural rather than oracular: an informed upstream restriction of the search space is a better inheritance than uncalibrated breadth, precisely the model's claim. Second, the same machine step creates value in one seat and merely salvages it in the other. Downstream of the expert, regularization is a fine-tuner, hunting for functional-form interdependencies among variables already known to matter; downstream of machine breadth, it becomes a rescue filter that recovers roughly a third of the penalty the novice configuration pays yet does not close the gap to the expert-first architecture. Selection after the fact cannot fully undo the absence of judgment before the search, and where judgment is absent at both ends the wastage arrives at the prediction stage intact.

---------- Insert Figure 13 about here ----------

The demonstration is deliberately bounded: it validates the architectural ordering the model predicts, in one empirical domain with one estimator family, with nine of ten pairwise comparisons favoring the expert arm and three reaching conventional significance. One correspondence is looser than the others: in the model, Case 3 pairs a low-performing upstream human with untethered, memoryless machine exploration, whereas its empirical counterpart here is defined architecturally, as the absence of informed judgment at either end of the machine's search. More generally, the experiment does not test the underlying memory mechanisms, which operate at a finer grain than a specification-choice experiment can resolve. The simulation supplies the mechanism; the experiment shows its footprint in a workflow any empirical researcher will recognize. The complete design, concept mapping, tables, and figures are reported in the *Supplemental Appendix (can be shared upon request to the authors)*.

### 6.4. Boundary Conditions and Limitations

The model's claims are bounded by its stylizations, and we state the principal ones plainly. *Archetypes, not descriptions.* Human adaptation is not exhausted by recency weighting-satisficing with aspiration levels, lexicographic rules, similarity-based search, and domain routines are all live alternatives (Gigerenzer and Gaissmaier 2011, Artinger et al. 2022)-and contemporary AI systems do not update by windowed equal-weight averaging. Our claims are about the interaction of memory regimes with task structure; the association of the regimes with human and artificial agents is a motivated interpretation, not an identity. A reader who prefers to read the model as contrasting more and less recency-biased searchers loses none of the structural results, though the parameter ranges (in particular, the machine agent's far larger scope and the feasibility of randomization at scale) are what tie those results to the human-AI application. Relatedly, the model gives both agents uniform capability across their sequences, whereas real AI exhibits a jagged frontier-strong in some regions of a task and weak in adjacent ones (Dell'Acqua et al. 2023). Introducing region-dependent machine reliability would transform the design problem into a dynamic delegation problem-when should the human act rather than delegate-which we regard as the most valuable extension of this framework.

*No cost of search.* Search is free in the model, which biases the results toward breadth; with realistic costs of generating and evaluating candidates-where AI's advantage is precisely its lower cost per evaluation (Csaszar et al. 2024)-the very-large-scope results of Sections 4.1 and 4.3 should be read as upper bounds, and the wastage of Figure 8 as a lower bound on the economic case against machine-breadth-before-human-filter architectures.

*Additive payoffs and conservative complementarity.* Because joint payoff averages the agents' payoffs, the model cannot represent super-additive synergy; our complementarity claims are strictly of the conservative, architecture-comparison form defined in Section 2.1, and for sequenced structures the pure-regime counterfactual battery (H-H and AI-AI in both orders) remains to be built. Embedding the memory-regime contrast in payoff structures with genuine cross-agent interaction-the coupled landscapes of Ethiraj and Levinthal (2004) and the patterned interdependence designs of Rivkin and Siggelkow (2003), with one decision maker replaced by a uniform-memory agent-is the natural next model in this program and would also restore the canonical epistasis that our sequential-adaptation implementation deliberately set aside. Our present results speak to sequential coupling in decision processes, not to landscape ruggedness, and we have confined our language accordingly (Section 3.5).

*Minimal agency.* The machine agent here follows a fixed updating rule; it does not form goals, orchestrate tools, or revise preferences-capacities that current scholarship treats as constitutive of agentic AI (Vanneste and Puranam 2024, Ide and Talamas 2025). We have therefore avoided framing the model as a theory of agentic AI as such: it is a theory of task architecture for agents with contrasting memory regimes, one pole of which is motivated by algorithmic consistency. Similarly, our memoryless regime is untethered exploration, not hallucination; modeling hallucination properly-as confident, systematic, non-random error, plausibly an *attractor-biasing* rather than attractor-escaping force-would likely reverse the benign role randomness plays here and is an important open question.

### 6.5. Conclusion

How organizations should divide and sequence decision-making between humans and machines is a design question, and design questions require mechanisms. This paper contributes a deliberately transparent one: two archetypal memory regimes-fast, recency-weighted adaptation and slow, uniform-memory adaptation-interacting over modular and sequenced task structures, with every result traceable to the threshold dynamics of adaptation. The headline finding inverts a prevailing template: machine search creates the most value not when it precedes human judgment, but when it follows high-quality human judgment, compounding what it inherits; and when what it would inherit is poor, untethered breadth beats faithful refinement. The experiment of Section 6.3 gives the ordering an empirical face: in predicting M&A announcement returns across time, expert-initiated search followed by machine fine-tuning transferred best, machine breadth filtered downstream came second, and a novice configuration in which the machine did all the work came last, reproducing the expected ordering of the model's three canonical cases. If the conclusion has a single sentence, it is this: in human-AI collaboration, the machine's defining behavioral property is that it amplifies its inheritance-so the central act of organization design is choosing what it inherits.

## References

Acar OA (2023) AI prompt engineering isn't the future. Harvard Bus. Rev. (June 6), https://hbr.org/2023/06/ai-prompt-engineering-isnt-the-future.

Acemoglu D, Restrepo P (2019) Automation and new tasks: How technology displaces and reinstates labor. J. Econom. Perspect. 33(2):3-30.

Agrawal A, Gans JS, Goldfarb A (2019) Exploring the impact of artificial intelligence: Prediction versus judgment. Inform. Econom. Policy 47:1-6.

Artinger FM, Gigerenzer G, Jacobs P (2022) Satisficing: Integrating two traditions. J. Econom. Literature 60(2):598-635.

Baumann O, Siggelkow N (2013) Dealing with complexity: Integrated vs. chunky search processes. Organ. Sci. 24(1):116-132.

Billinger S, Srikanth K, Stieglitz N, Schumacher TR (2021) Exploration and exploitation in complex search tasks: How feedback influences whether and where human agents search. Strategic Management J. 42(2):361-385.

Choudhary V, Marchetti A, Shrestha YR, Puranam P (2025) Human-AI ensembles: When can they work? J. Management 51(2):536-569.

Choudhury P, Starr E, Agarwal R (2020) Machine learning and human capital complementarities: Experimental evidence on bias mitigation. Strategic Management J. 41(8):1381-1411.

Christensen M, Knudsen T (2013) How decisions can be organized-and why it matters. J. Organ. Design 2(3):41-50.

Clement J, Puranam P (2018) Searching for structure: Formal organization design as a guide to network evolution. Management Sci. 64(8):3879-3895.

Csaszar FA (2025) Artificial intelligence and strategic decision-making. Working paper, University of Michigan, Ann Arbor.

Csaszar FA, Eggers JP (2013) Organizational decision making: An information aggregation view. Management Sci. 59(10):2257-2277.

Csaszar FA, Ketkar H, Kim H (2024) Artificial intelligence and strategic decision-making: Evidence from entrepreneurs and investors. Strategy Sci. 9(4):322-345.

Cyert RM, March JG (1963) A Behavioral Theory of the Firm (Prentice-Hall, Englewood Cliffs, NJ).

Dell'Acqua F, McFowland E III, Mollick ER, Lifshitz-Assaf H, Kellogg K, Rajendran S, Krayer L, Candelon F, Lakhani KR (2023) Navigating the jagged technological frontier: Field experimental evidence of the effects of AI on knowledge worker productivity and quality. Working Paper 24-013, Harvard Business School, Boston.

Ethiraj SK, Levinthal D (2004) Bounded rationality and the search for organizational architecture: An evolutionary perspective on the design of organizations and their evolvability. Admin. Sci. Quart. 49(3):404-437.

Ganco M, Kapoor R, Lee GK (2020) From rugged landscapes to rugged ecosystems: Structure of interdependencies and firms' innovative search. Acad. Management Rev. 45(3):646-674.

Gigerenzer G, Gaissmaier W (2011) Heuristic decision making. Annual Rev. Psych. 62:451-482.

Gigerenzer G, Goldstein DG (1996) Reasoning the fast and frugal way: Models of bounded rationality. Psych. Rev. 103(4):650-669.

Greve HR (2003) Organizational Learning from Performance Feedback: A Behavioral Perspective on Innovation and Change (Cambridge University Press, Cambridge, UK).

Harrison JR, Lin Z, Carroll GR, Carley KM (2007) Simulation modeling in organizational and management research. Acad. Management Rev. 32(4):1229-1245.
Ide E, Talamas E (2025) Artificial intelligence in the knowledge economy. J. Political Econom., ePub ahead of print, https://www.journals.uchicago.edu/doi/10.1086/737233.
Kauffman SA (1993) The Origins of Order: Self-Organization and Selection in Evolution (Oxford University Press, New York).
Knudsen T, Levinthal DA, Puranam P (2019) Editorial: A model is a model. Strategy Sci. 4(1):1-3.
Knudsen T, Srikanth K (2014) Coordinated exploration: Organizing joint search by multiple specialists to overcome mutual confusion and joint myopia. Admin. Sci. Quart. 59(3):409-441.
Krakowski S, Luger J, Raisch S (2023) Artificial intelligence and the changing sources of competitive advantage. Strategic Management J. 44(6):1425-1452.
Levinthal DA (1997) Adaptation on rugged landscapes. Management Sci. 43(7):934-950.
Levinthal DA, March JG (1981) A model of adaptive organizational search. J. Econom. Behav. Organ. 2(4):307-333.
Levinthal DA, Schliesmann D (2025) Cautious exploitation: Learning and search in problems of evaluation and discovery. Organ. Sci. 36(2):903-917.
March JG (1991) Exploration and exploitation in organizational learning. Organ. Sci. 2(1):71-87.
Nishant R, Schneckenberg D, Ravishankar MN (2024) The formal rationality of artificial intelligence-based algorithms and the problem of bias. J. Inform. Tech. 39(1):19-40.
Pearl J (1984) Heuristics: Intelligent Search Strategies for Computer Problem Solving (Addison-Wesley, Reading, MA).
Posen HE, Keil T, Kim S, Meissner FD (2018) Renewing research on problemistic search-A review and research agenda. Acad. Management Ann. 12(1):208-251.
Puranam P (2021) Human-AI collaborative decision-making as an organization design problem. J. Organ. Design 10(2):75-80.
Puranam P, Stieglitz N, Osman M, Pillutla MM (2015) Modelling bounded rationality in organizations: Progress and prospects. Acad. Management Ann. 9(1):337-392.
Raisch S, Fomina K (2025) Combining human and artificial intelligence: Hybrid problem-solving in organizations. Acad. Management Rev. 50(2):441-464.
Raisch S, Krakowski S (2021) Artificial intelligence and management: The automation-augmentation paradox. Acad. Management Rev. 46(1):192-210.
Rivkin JW, Siggelkow N (2003) Balancing search and stability: Interdependencies among elements of organizational design. Management Sci. 49(3):290-311.
Russell SJ, Norvig P (2010) Artificial Intelligence: A Modern Approach, 3rd ed. (Prentice-Hall, Upper Saddle River, NJ).
Shick M, Johnson N, Fan Y (2024) Artificial intelligence and the end of bounded rationality: A new era in organizational decision making. Development Learning Organ. 38(4):1-3.
Simon HA (1955) A behavioral model of rational choice. Quart. J. Econom. 69(1):99-118.
Simon HA (2000) Bounded rationality in social science: Today and tomorrow. Mind Soc. 1(1):25-39.
Simon HA, Newell A (1958) Heuristic problem solving: The next advance in operations research. Oper. Res. 6(1):1-10.
Thompson JD (1967) Organizations in Action: Social Science Bases of Administrative Theory (McGraw-Hill, New York).

Todd PM, Gigerenzer G (2007) Environments that make us smart: Ecological rationality. Current Directions Psych. Sci. 16(3):167-171.

Vanneste BS, Puranam P (2024) Artificial intelligence, trust, and perceptions of agency. Acad. Management Rev., ePub ahead of print.

## Tables & Figures

### Table 1. Model Constructs and their Relationship to Canonical NK/NKC Counterparts

| Construct | Role in this model | Canonical NK/NKC counterpart | Key departure |
|---|---|---|---|
| $N_a$ (task scope) | Number of binary sub-decisions in agent *a*'s task; length of its decision sequence. Scope ratio $N_{AI}/N_H > 1$ is a design variable. | Dimensionality of the decision space over which agents reconfigure. | Sequence length in a progressive-realization process, not the dimension of a combinatorial configuration space. |
| $K_a$ (coupling depth) | Number of preceding realized decisions conditioning agent *a*'s next decision; depth of within-task path dependence. | Epistatic interdependence among decisions in the payoff (fitness) function; source of landscape ruggedness. | Enters the adaptation rule, not the payoff function; a process property (history window), not environmental epistasis. |
| *C* (cross-agent coupling) | Number of upstream realized decisions inherited as the downstream agent's initial history (C = 0: modular). | Coupling of payoff functions across co-evolving landscapes (NKC). | Inherited history rather than cross-landscape epistasis; governs intensity of sequential interdependence. |
| Payoff | Share of sub-decisions resolved to 1; joint payoff (APO) averages the two agents' payoffs. | Fitness averaged over interdependent contribution functions; rugged, multi-peaked surface. | Additive; unique global optimum (all ones); no epistasis. Dynamics, not topology, generate trap-like behavior. |
| Search process | Progressive realization: each sub-decision is decided once, in order, from a weighted window of realized history. | Movement between complete configurations via neighborhood (local) or distant search. | No reconfiguration or revisiting; the local/distant search distinction has no direct counterpart. |
| "Local trap" analogue | Absorbing low-payoff regime: a trajectory whose | Local peak: configuration whose | A dynamic property of adaptation, not a |

| Construct | Role in this model | Canonical NK/NKC counterpart | Key departure |
|---|---|---|---|
| | window mean sits below the threshold reproduces failures (Section 3.6). | one-bit neighbors are all inferior. | topological property of the payoff surface. |
| Agent regimes | Recency-weighted (human archetype); uniform-memory (AI archetype); memoryless stochastic (untethered exploration). | Typically homogeneous boundedly rational local searchers. | Heterogeneous memory regimes are the object of study; all regimes are heuristics (Section 2.2). |

Note. The model retains N-K-C notation to preserve comparability of design questions with the NK/NKC tradition; the mechanics differ as stated. Readers may equivalently read the model as a bespoke sequential-adaptation model (Section 3.5).

**Table 2. Worked Example: Inheritance and Memory Fade in the H→AI Structure**

| Element | Conditioning window (oldest → most recent) | Source | Window mean | Decision |
|---|---|---|---|---|
| $x_1^{AI}$ | (1, 0, 0, 1) | Inherited ($S^{AI}$) | 0.50 | 1 |
| $x_2^{AI}$ | (0, 0, 1, 0) | Inherited ($S^{AI}$) | 0.25 | 0 |
| $x_3^{AI}$ | (0, 1, 0, 1) | Inherited ($S^{AI}$) | 0.50 | 1 |
| $x_4^{AI}$ | (1, 0, 1, $x_1^{AI}$ = 1) | Inherited + own states | 0.75 | 1 |
| $x_5^{AI}$ | (0, 1, $x_1^{AI}$ = 1, $x_2^{AI}$ = 0) | Inherited + own states | 0.50 | 1 |
| $x_6^{AI}$ | (1, $x_1^{AI}$ = 1, $x_2^{AI}$ = 0, $x_3^{AI}$ = 1) | Inherited + own states | 0.75 | 1 |
| $x_7^{AI}$ | ($x_3^{AI}$, $x_4^{AI}$, $x_5^{AI}$, $x_6^{AI}$) = (1, 1, 1, 1) | Own states only (memory faded) | 1.00 | 1 |
| $x_8^{AI}$ | ($x_4^{AI}$, $x_5^{AI}$, $x_6^{AI}$, $x_7^{AI}$) = (1, 1, 1, 1) | Own states only (memory faded) | 1.00 | 1 |

Note. Parameters: $N_h$ = 6, C = 6, $N_{ai}$ = 8, $K_{ai}$ = 4; inherited human states $S^{AI}$ = (1, 0, 0, 1, 0, 1); downstream regime: uniform-memory adaptation; threshold ties (window mean = 0.50) resolve to 1. From $x_7^{AI}$ onward the rolling window contains only machine-generated states-the inherited human search has faded from memory. Under the memoryless downstream regime, decisions after the inherited segment are instead drawn Bernoulli(0.5).

## Figures

### Figure 1. Baseline Payoff Under Pure Adaptation Regimes

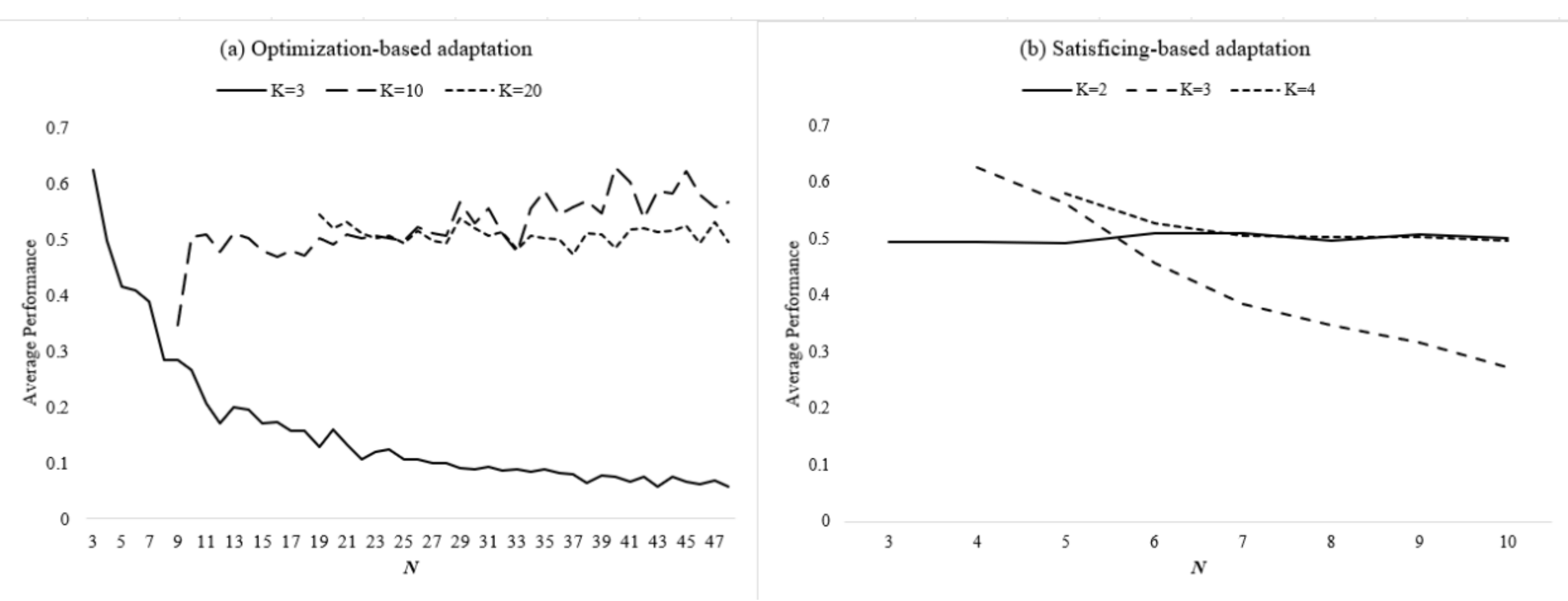

Note. Single-regime baselines across (N, K): (a) uniform-memory adaptation; (b) recency-weighted adaptation. Means over 1,000 runs per cell; cubic-spline smoothing. These surfaces double as exact pure-regime counterfactuals for the modular analysis (Section 3.6).

### Figure 2. Worked Example: The Two Memory Regimes Reach Different Decisions on an Identical History

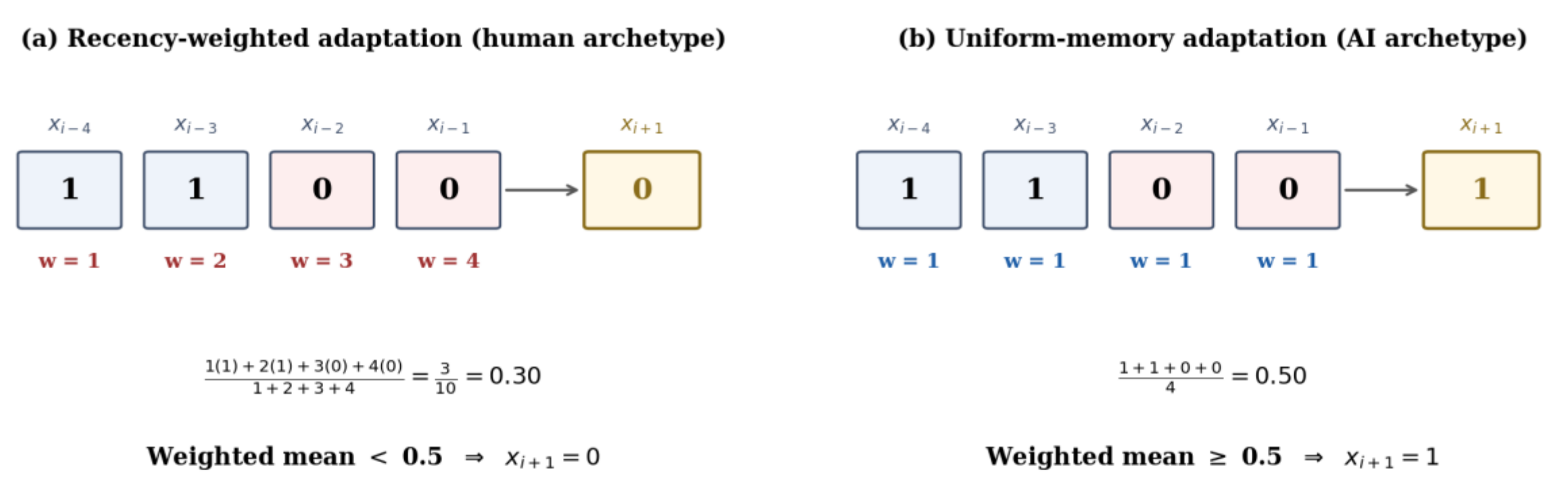

Note. Window K = 4; realized history (1, 1, 0, 0), oldest to most recent. Recency weights (1, 2, 3, 4) yield a weighted mean of 0.30 → decision 0; uniform weights yield 0.50 → decision 1 (threshold ties resolve to 1).

## Figure 3. Joint Payoff Under Modular Task Structures (C = 0)

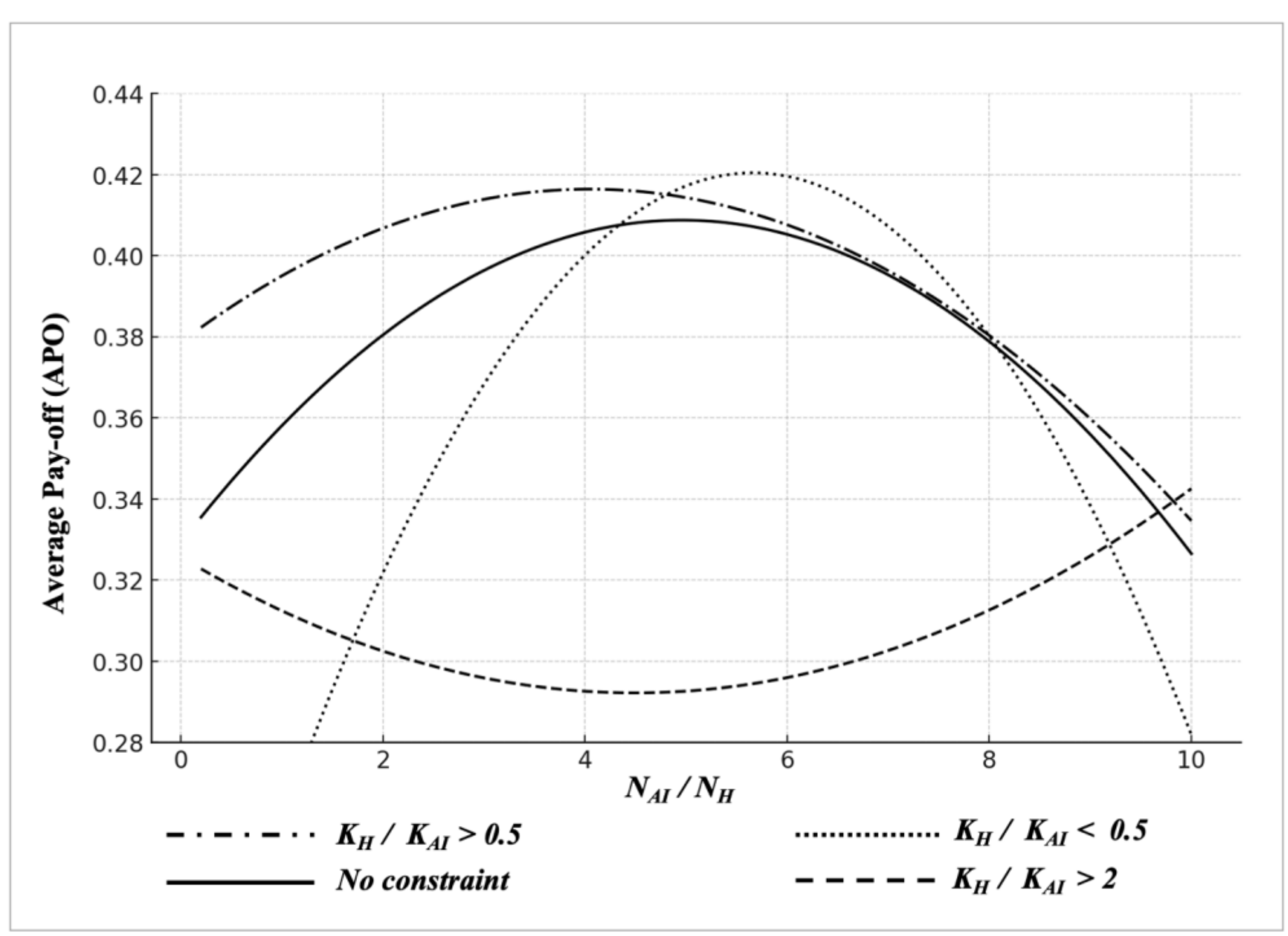

Note. Joint payoff (APO) against relative scope $N_{ai}/N_h$ for varying coupling-depth ratios $K_h/K_{ai}$. Means over 1,000 runs per parameter combination; cubic-spline smoothing.

## Figure 4. Joint Payoff Under the AI→H Sequenced Structure (C > 0)

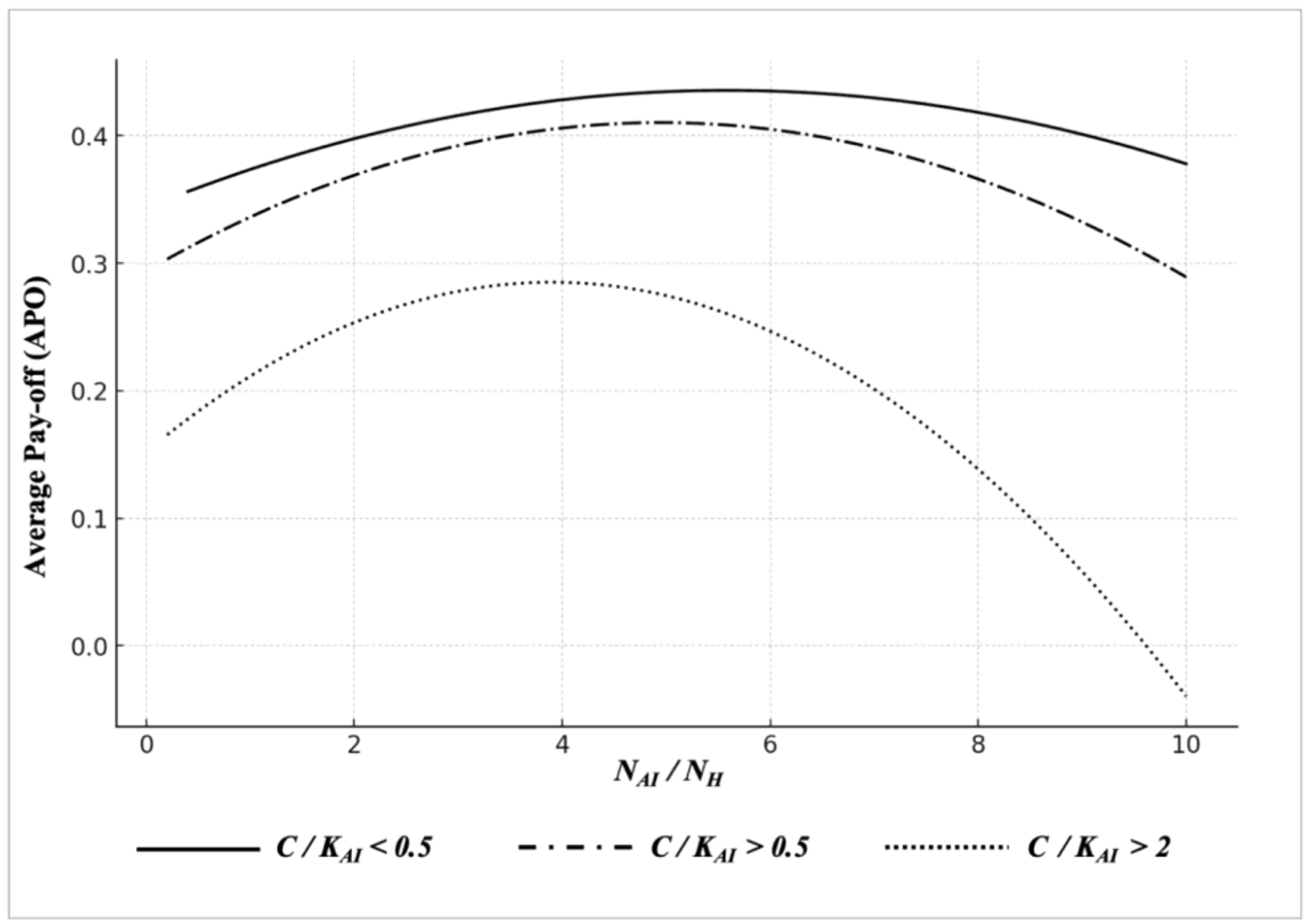

Note. Joint payoff against relative scope for varying cross-agent coupling; means over 1,000 runs per combination; cubic-spline smoothing.

**Figure 5. Joint Payoff Under the H→AI Sequenced Structure (C > 0)**

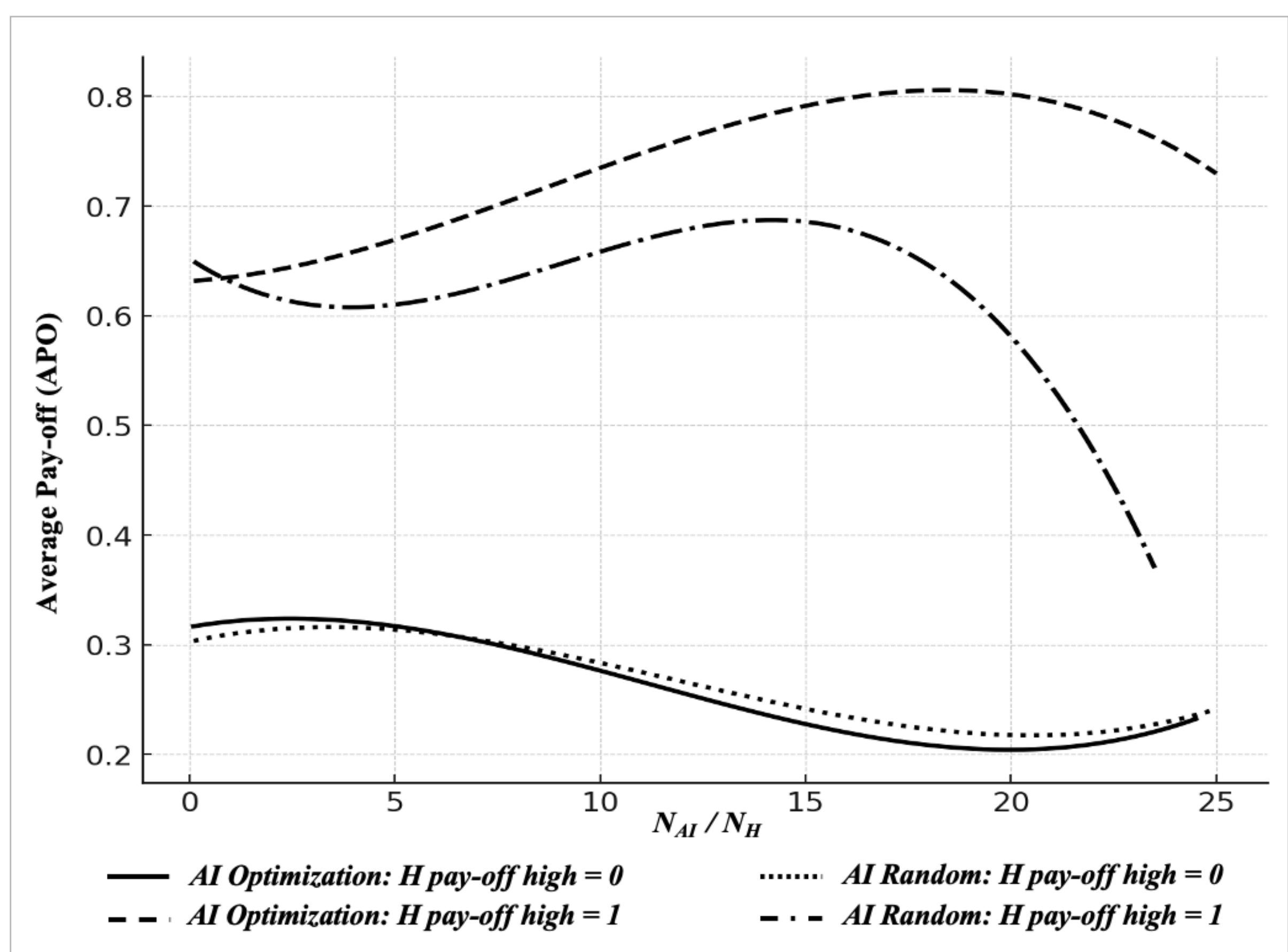

Note. Downstream regimes: uniform-memory and memoryless stochastic adaptation, by quality of the upstream human trajectory. Means over 1,000 runs per combination; cubic-spline smoothing.

## Figure 6. Sequence Comparison: AI→H versus H→AI

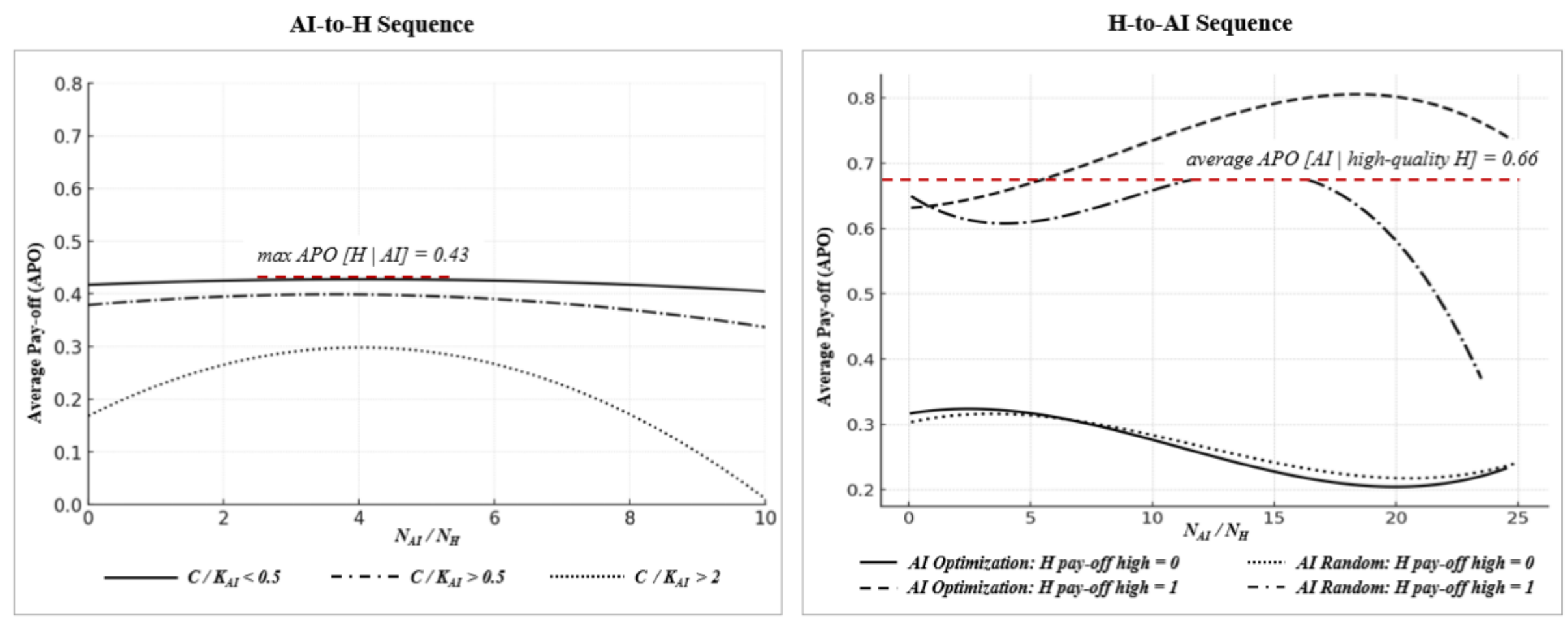

Note. The H→AI structure seeded by a high-performing upstream human dominates: its average joint payoff exceeds the maximum of the AI→H structure. Cubic-spline smoothing.

## Figure 7. Three Canonical Cases of Sequenced Human-AI Search

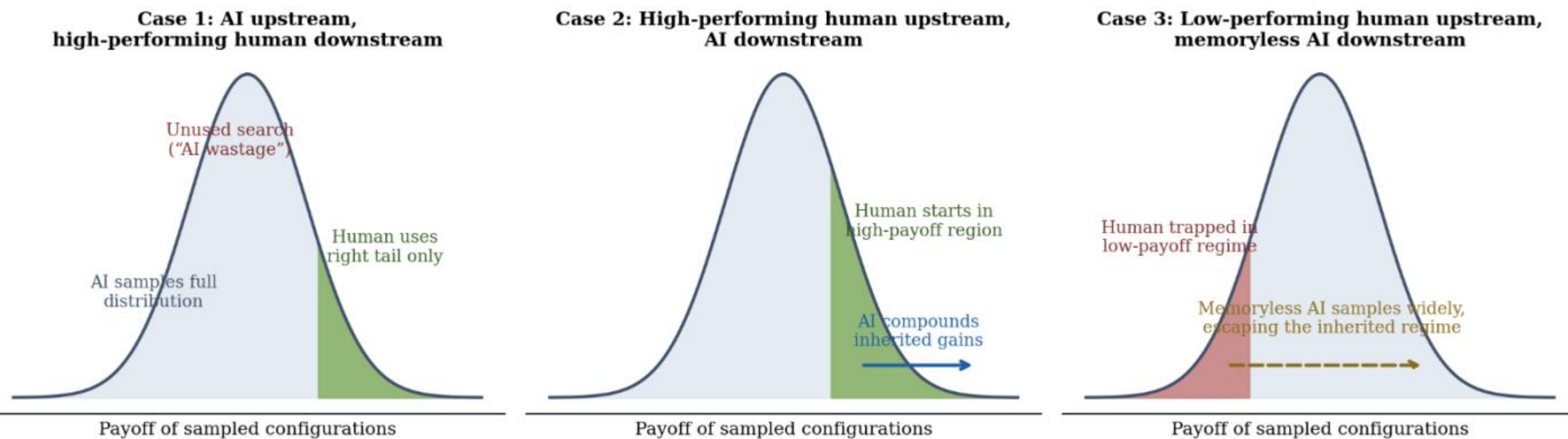

Note. Conceptual schematic. Case 1: broad upstream machine search, thinly utilized downstream (search wastage). Case 2: high-quality upstream human search, compounded by downstream uniform-memory adaptation. Case 3: low-quality upstream human search, escaped by downstream memoryless breadth. Expected performance: Case 2 > Case 1 > Case 3.

## Figure 8. Search Wastage in the AI→H Structure

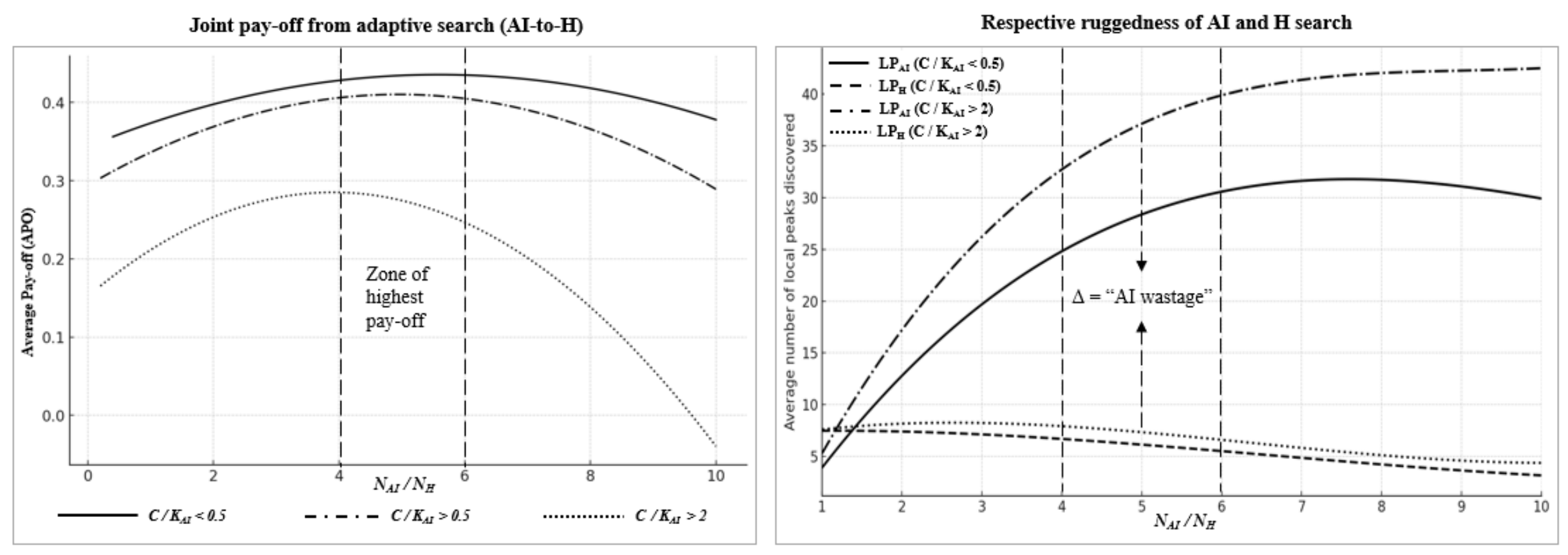

Note. Final payoff against interim solutions generated by upstream machine search and utilized by downstream human adaptation (operational definition in Section 4.4). Within the high-payoff scope band ($4 < N_{ai}/N_h < 6$), upstream search generates over 3.5 times more interim solutions than downstream adaptation utilizes.

## Figure 9. Sensitivity of Modular Performance to Scope and Coupling Depth

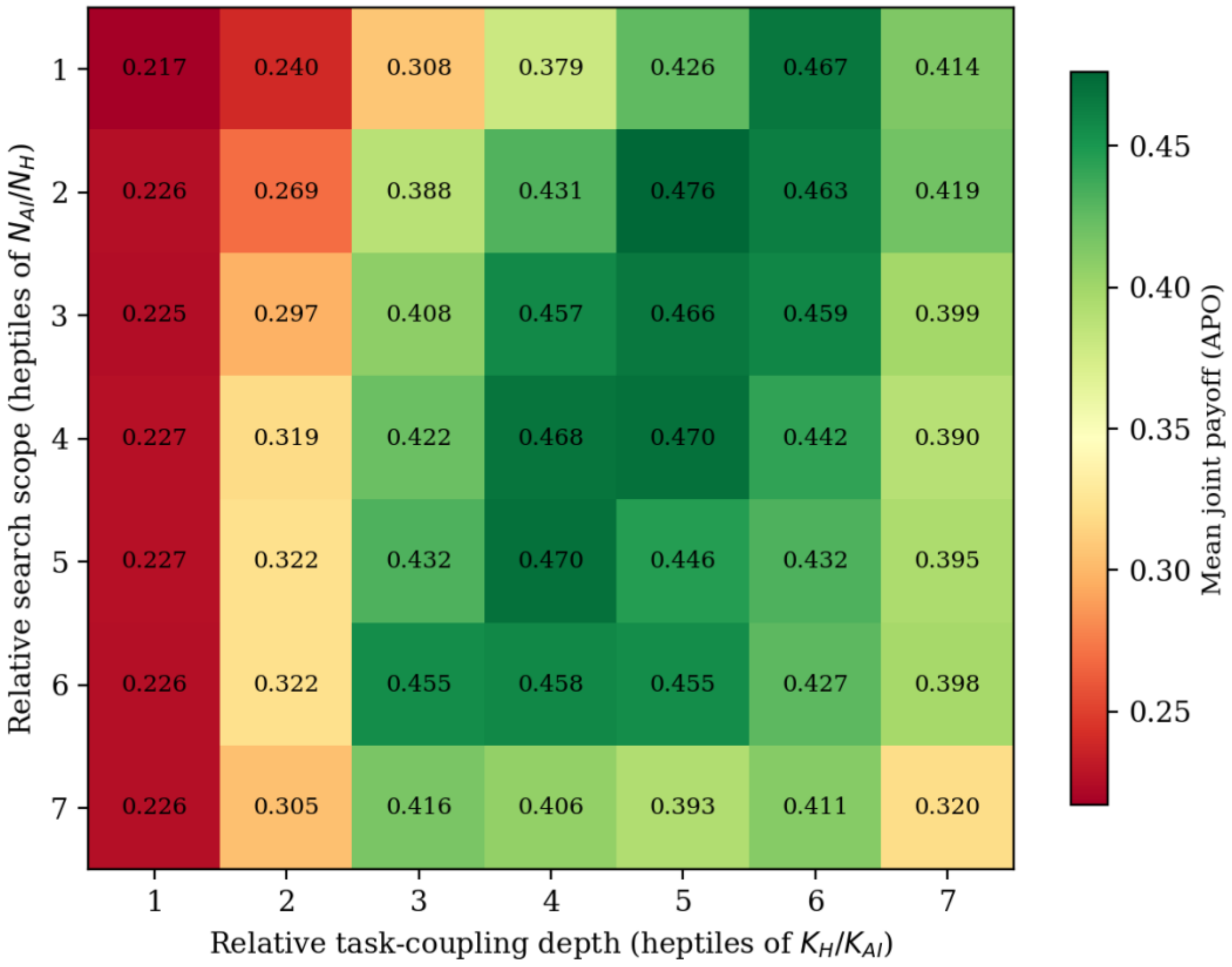

Note. Unsmoothed mean joint payoff (1,000 runs per cell). Rows: heptiles of relative scope $N_{ai}/N_h$; columns: heptiles of relative coupling depth $K_h/K_{ai}$.

**Figure 10. Sensitivity of AI→H Performance to Within-Task versus Cross-Agent Coupling**

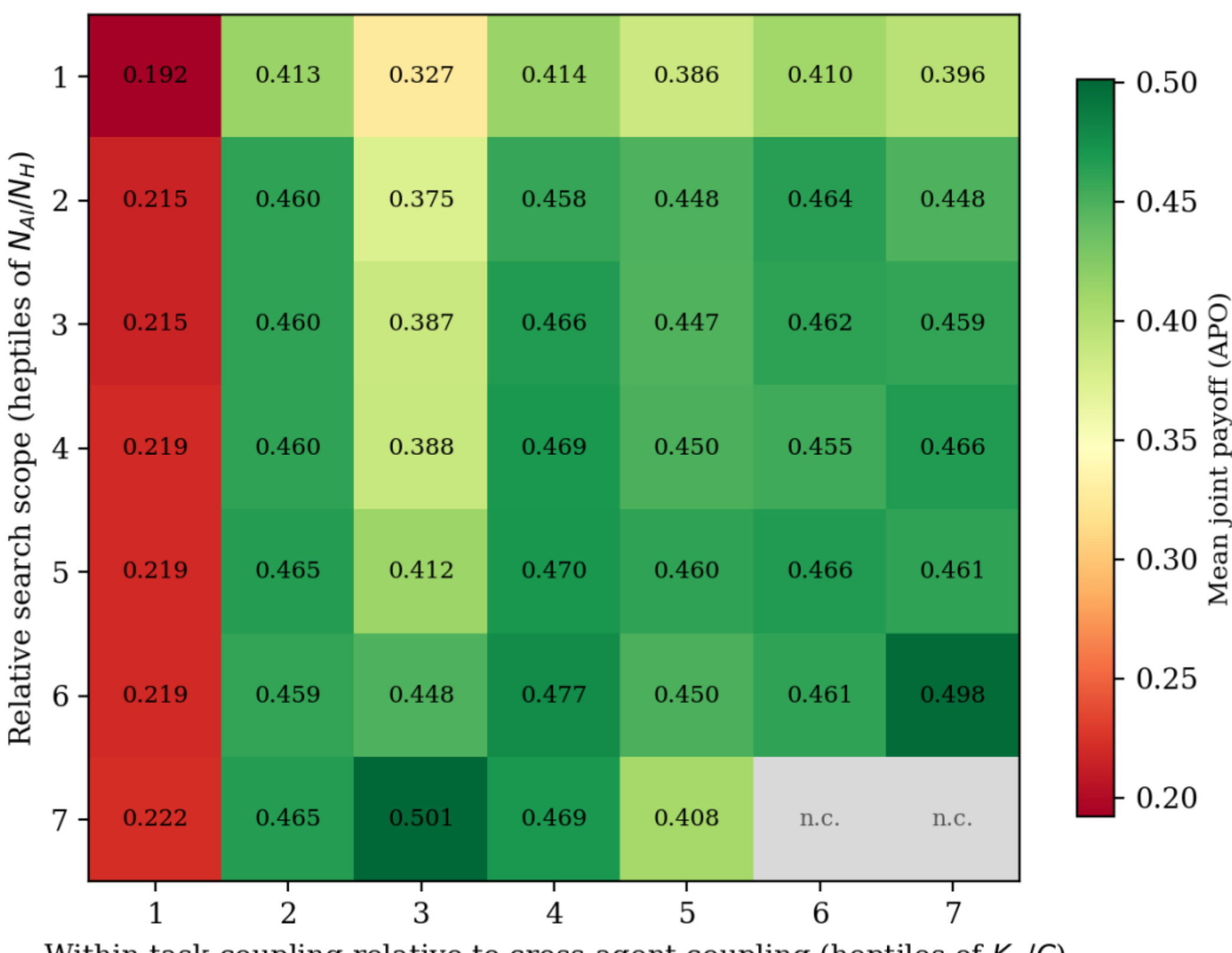

Note. Unsmoothed mean joint payoff (1,000 runs per cell). Rows: heptiles of relative scope; columns: heptiles of $K_h/C$. "n.c." marks cells that failed to converge within the run budget.

## Figure 11. Sensitivity of H→AI Performance to Cross-Agent Coupling

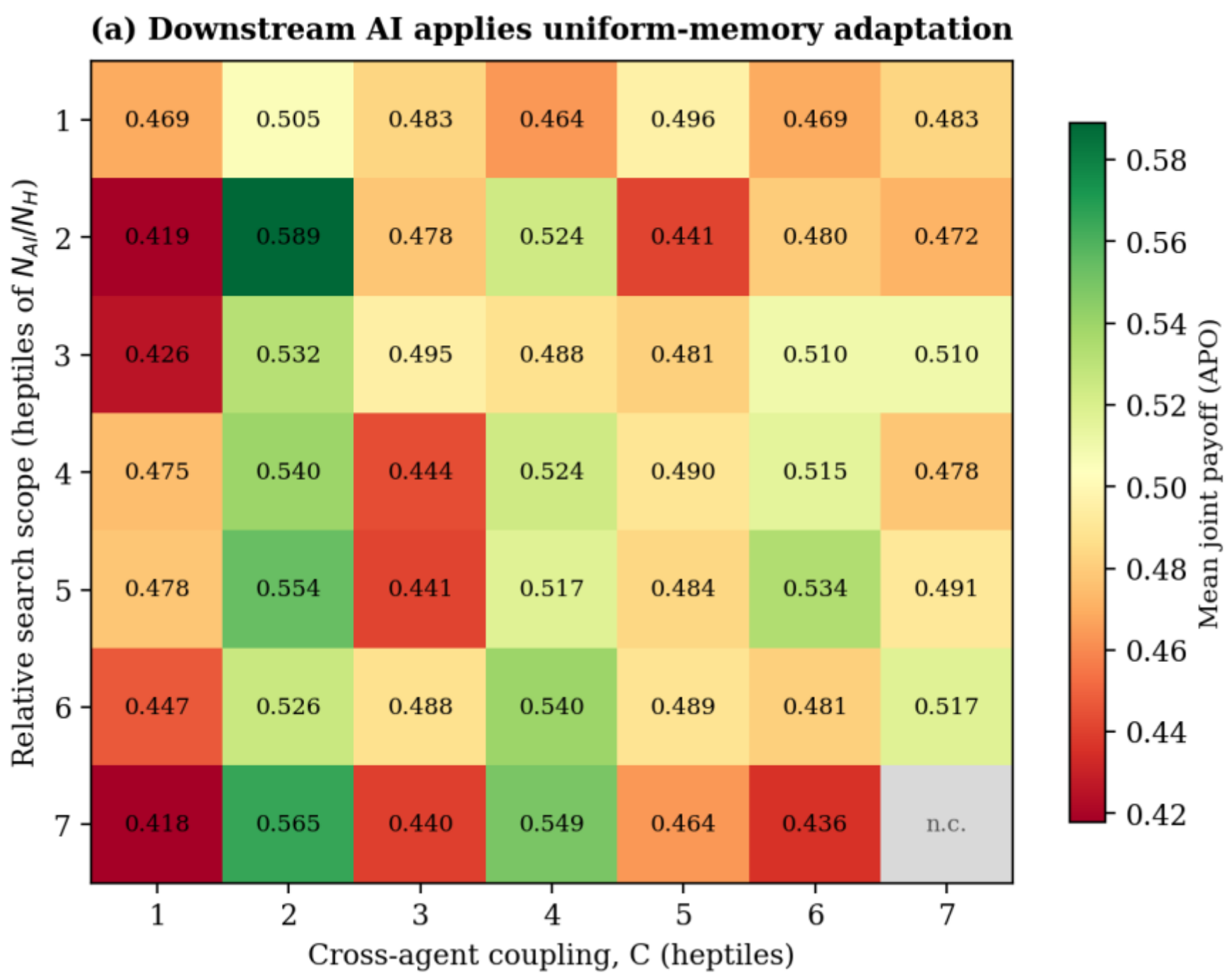

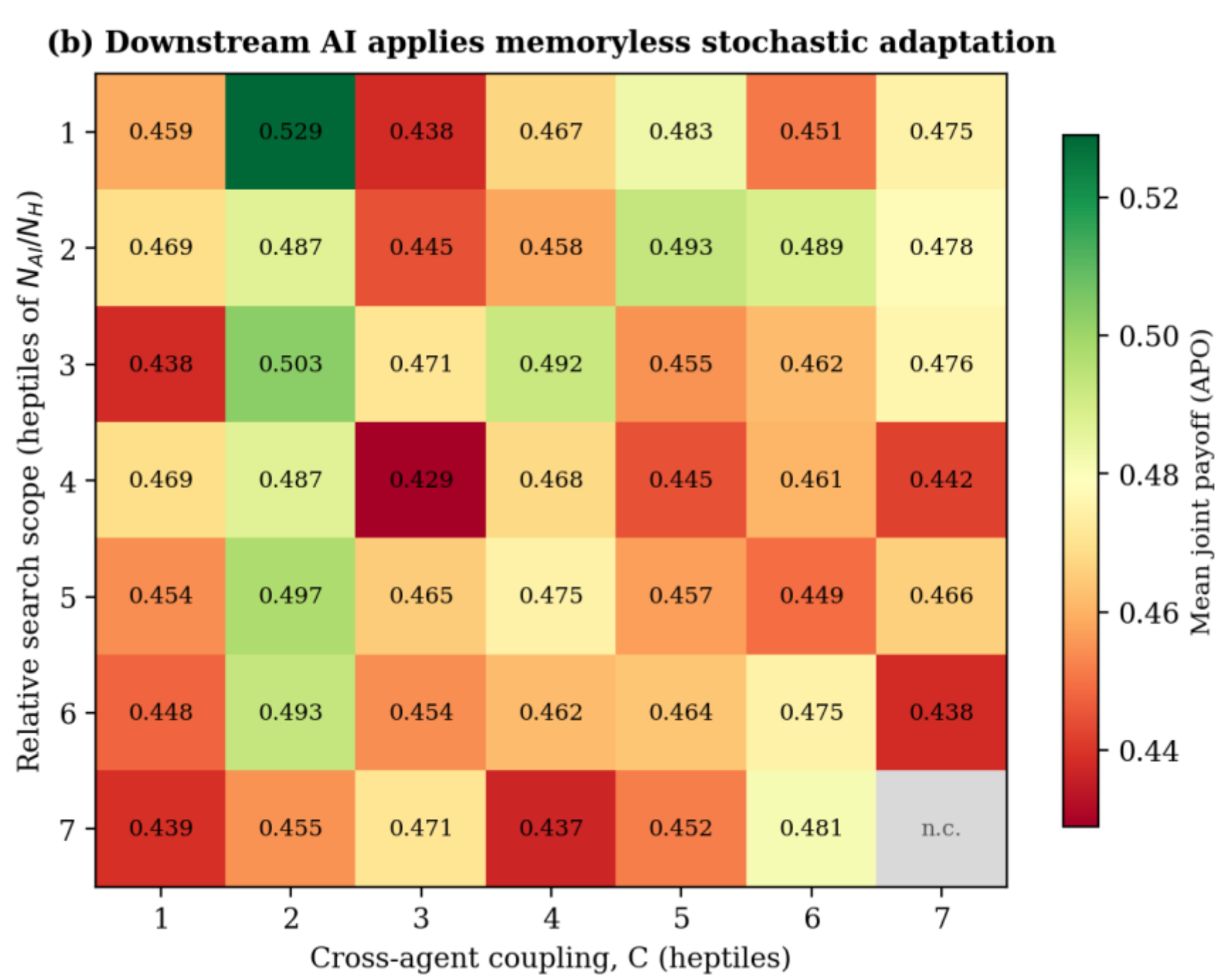

Note. Unsmoothed mean joint payoff (1,000 runs per cell). Rows: heptiles of relative scope; columns: heptiles of cross-agent coupling C. Panel (a): downstream uniform-memory adaptation; panel (b): downstream memoryless stochastic adaptation. “n.c.” marks non-converged cells.

**Figure 12. The Three Search Architectures of the Experimental Validation**

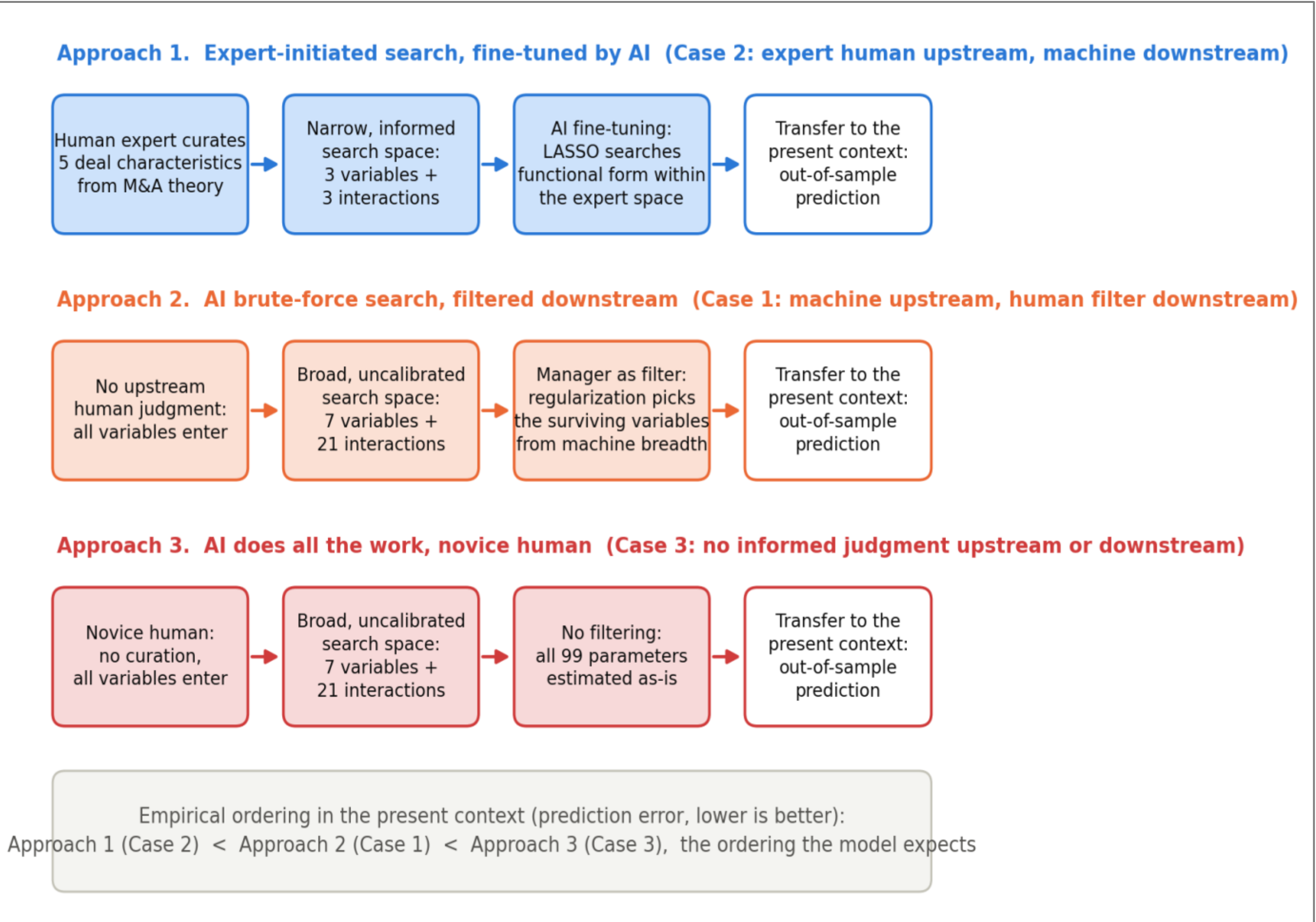

Note. Approach 1 instantiates Case 2 of the model (expert human upstream, machine downstream): an expert curates five theory-relevant deal characteristics, three enter at a time, and a penalized learner (LASSO) fine-tunes the functional form, the linear terms and their pairwise interactions, within that informed space. Approach 2 instantiates Case 1 (machine upstream, human filter downstream): all seven variables and all 21 pairwise interactions enter, and the regularization step acts as a downstream filter, a proxy for managers picking the surviving variables. Approach 3 instantiates Case 3 (novice human, no informed judgment at either end): the machine does all the work, and the full 99-parameter specification is estimated as-is. In all arms the selected terms are re-estimated by OLS with industry fixed effects on past deals (2008 to 2013) and predict announcement returns for present deals (2014 to 2020).

**Figure 13. Experimental Validation: Transfer to the Present Context**

The empirical ordering matches the model: Case 2 above Case 1 above Case 3

naive baseline
(predict past mean)

expert-initiated models (10)
Case 2

novice: unfiltered breadth
Case 3

machine breadth, filtered
Case 1

0.07630 0.07635 0.07640 0.07645

Out-of-sample RMSE in the present context, Sample 2 (lower is better)

Note. Out-of-sample RMSE predicting CAR for present deals (Sample 2; N = 16,930; lower is better). Blue dots: the ten expert-initiated models, Case 2 (jittered vertically). Orange diamond: filtered machine breadth, Case 1 (LASSO selection over 28 candidate terms). Red cross: the novice configuration, Case 3 (unfiltered 99-parameter OLS). Dashed line: naive forecast of the past mean. Nine of ten expert-initiated models beat the filtered brute-force model and all ten beat the novice configuration; no specification beats the naive benchmark, so the expert-first architecture wins by losing less. The empirical ordering reproduces the expected ordering of the model's three canonical cases. Full design and results in the Online Appendix.